\documentclass[preprint2]{aastex63}

\def\gsim{\mathrel{\rlap{\lower 4pt \hbox{\hskip 1pt $\sim$}}\raise 1pt
\hbox {$>$}}}
\def\lsim{\mathrel{\rlap{\lower 4pt \hbox{\hskip 1pt $\sim$}}\raise 1pt
\hbox {$<$}}}

\received{}
\revised{}
\accepted{by ApJ on 25 October, 2022}
\shorttitle{SN Iax 2019muj in the late phase}
\shortauthors{Maeda et al.}
\graphicspath{{./}{figures/}}

\begin{document}

\title{Properties of Type Iax Supernova 2019muj in the Late Phase: \\
Existence, Nature and Origin of the Iron-rich Dense Core}

\correspondingauthor{Keiichi Maeda}
\email{keiichi.maeda@kusastro.kyoto-u.ac.jp}

\author[0000-0003-2611-7269]{Keiichi Maeda}
\affiliation{Department of Astronomy, Kyoto University, Kitashirakawa-Oiwake-cho, Sakyo-ku, Kyoto, 606-8502. Japan}

\author[0000-0002-4540-4928]{Miho Kawabata}
\affiliation{Department of Astronomy, Kyoto University, Kitashirakawa-Oiwake-cho, Sakyo-ku, Kyoto, 606-8502. Japan}



\begin{abstract}
Type Iax Supernovae (SNe Iax) form a class of peculiar SNe Ia, whose early-phase spectra share main spectral line identifications with canonical SNe Ia but with higher ionization and much lower line velocities. Their late-time behaviors deviate from usual SNe Ia in many respects; SNe Iax keep showing photospheric spectra over several 100 days and the luminosity decline is very slow. In the present work, we study the late-time spectra of SN Iax 2019muj including a newly-presented spectrum at $\sim 500$ days. The spectrum is still dominated by allowed transitions but with lower ionization state, with possible detection of [O I]$\lambda\lambda$6300, 6363. By comprehensively examining the spectral formation processes of allowed transitions (Fe II, Fe I, and the Ca II NIR triplet) and forbidden transitions ([Ca II]$\lambda\lambda$7292, 7324 and the [O I]), we quantitatively constrain the nature of the innermost region and find that it is distinct from the outer ejecta; the mass of the innermost component is $\sim 0.03 M_\odot$ dominated by Fe (which can be initially $^{56}$Ni), expanding with the velocity of $\sim 760$ km s$^{-1}$. We argue that the nature of the inner component is explained by the failed/weak white-dwarf thermonuclear explosion scenario. We suggest that a fraction of the $^{56}$Ni-rich materials initially confined in (the envelope of) the bound remnant can later be ejected by the energy input through the $^{56}$Ni/Co/Fe decay, forming the `second' unbound ejecta component which manifests itself as the inner dense component seen in the late phase.
\end{abstract}



\section{Introduction} \label{sec:intro}

It has been widely accepted that Type Ia Supernovae (SNe Ia) are thermonuclear explosions of a C+O white dwarf (WD). Details of the progenitor system(s) and explosion mechanism(s) are however still under active debate \citep[e.g.,][for a review]{maeda2022}. There are two leading scenarios for the explosion mechanism(s). The delayed detonation scenario has been a leading scenario for many decades \citep{khokhlov1991,iwamoto1999,maeda2010,seitenzahl2013}, in which the thermonuclear runaway first triggers a deflagration flame near the center of a (nearly) Chandrasekhar-limiting mass ($M_{\rm Ch}$) WD, which later turns into the detonation wave through the deflagration-to-detonation transition (DDT). The double-detonation scenario is another popular scenario \citep{nomoto1982b,livne1990,woosley1994,shen2009}, which has attracted increasing attention from the community in the last decade. In this scenario, the He detonation triggered in a He envelope of a sub-$M_{\rm Ch}$ WD inserts the shock wave into a C+O WD core, and the second C detonation is triggered within the core, disrupting the whole WD. These explosion mechanisms are not exclusively associated with a specific progenitor system, i.e., the single-degenerate (SD) scenario \citep[i.e., a WD accreting materials from a non-degenerate companion:][]{whelan1973,nomoto1982a} or the double-degenerate (DD) scenario \citep[i.e., a binary WD merger:][]{iben1984,webbink1984}. For example, while the double-detonation mechanism was originally proposed within the context of the SD scenario, it has recently been suggested that the condition may also be met in the DD scenario \citep{guillochon2010,pakmor2013,tanikawa2015,shen2018}. 

It has not yet been clarified which (or anything else) is the `correct' combination of the explosion mechanism and the progenitor system leading to SNe Ia. Indeed, it is very likely that SNe Ia are a mixture of different populations with different progenitor systems and explosion mechanisms, given a huge diversity and emerging various subclasses seen in observational properties of SNe Ia. In a sense, one of the main goals in the field is to map SNe Ia originating in different progenitors and explosion mechanisms in the diversity zoo of observed SNe Ia \citep[e.g.,][]{maeda2016,taubenberger2017,jha2019}. 

A class of SNe Iax \citep{foley2013}, or SN 2002cx-like objects \citep{li2003,jha2006}, stands out as an extreme SN Ia subclass. Despite the spectral similarity between SNe Iax and normal SNe Ia, which share main spectral line identifications, there are some distinct observational properties seen in SNe Iax \citep[][for a review]{jha2017}; in the early phase in which they have been intensively observed, they tend to have much lower velocities in the spectral features ($\sim 2,000-8,000$ km s$^{-1}$ at the maximum light) and lower peak luminosities ($\sim -14$ to $-19$ magnitude) than normal SNe Ia ($\gsim 10,000$ km s$^{-1}$ and $\lsim -19$ mag). There is a general correlation between the early-phase light curve and spectral properties; brighter SNe Iax tend to evolve more slowly, and tend to have higher line velocities \citep{foley2013,magee2016} \citep[but see][]{magee2017}. 

These observational properties indicate that the masses of the ejecta and explosive-nucleosynthesis products (e.g., $^{56}$Ni), as well as the kinetic energy of the explosion, are smaller for SNe Iax than (canonical) SNe Ia \citep{stritzinger2014,szalai2015,yamanaka2015,kawabata2018,srivastav2020} \citep[but see][]{sahu2008,stritzinger2015}. This, together with the young environment found for many SNe Iax \citep{foley2009,lyman2018}, has raised a question as to whether they could be core-collapse SNe rather than thermonuclear SNe \citep{valenti2009,moriya2010}. The issue has not yet been completely settled, but there are mounting indications that point to the thermonuclear origin, including the spectral similarity to SNe Ia \citep{jha2017}, the detection of NIR Co II lines \citep{stritzinger2014,tomasella2016}, and the detection of a probable He star companion star in the pre-SN image of SN Iax 2012Z \citep{mccully2014a} that has not disappeared in the post-SN image \citep{mccully2022}. 

The argument for the thermonuclear origin has been further strengthened by several models in the context of the WD thermonuclear explosion scenario, which can account for major properties of SNe Iax. A currently popular scenario is the weak/failed deflagration model of a $M_{\rm ch}$ WD, which can be viewed as a variant of the delayed-detonation model but without the DDT \citep{jordan2012,kromer2013} \citep[but see][for other scenarios]{hoeflich1995,stritzinger2015}. The model predicts variations in the explosion energy and the ejected mass depending on the strength of the deflagration (which is controlled by the number of the ignition spots), leading to a range of expected observational (early-phase spectral and light-curve) properties covering the observational diversity \citep{fink2014,kromer2015,lach2022}. 
Spectral synthesis modeling efforts for individual SNe Iax show general agreement between the synthetic spectra based on unbound ejecta in this scenario and the early-phase spectra of SNe Iax \citep{magee2016,barna2017,barna2018,srivastav2020, barna2021}, while the details such as the expected composition mixing are still to be tested  \citep{magee2022}. A striking feature in the scenario is that a bound WD remnant can be left behind the explosion; if the nuclear energy generation is below $\sim 10^{51}$ erg, it is not sufficient to unbound all the progenitor WD \citep{fink2014}. We call this regime leaving the WD remnant the `failed' deflagration scenario. 

Despite the success of the failed/weak deflagration scenario, it is not the end of the story. It has been found that SNe Iax show distinct properties in the late phase ($\gsim 100$ days). Unlike normal SNe Ia, they never enter into the pure nebular phase but keep showing the photospheric spectra dominated by allowed transitions (mainly Fe II) \citep{jha2006,sahu2008,mccully2014b,foley2016}. 
This is qualitatively consistent with their light-curve properties, showing a slow evolution roughly following the $^{56}$Co decay with little leakage of the decay $\gamma$-rays (or even slower) \citep{foley2014,kawabata2018,kawabata2021,mccully2022}. These late-time properties point to the existence of a high-density, innermost region. It has been speculated that these properties may be associated with the putative WD remnant in the failed deflagration scenario \citep{foley2016,kawabata2021}, but the connection has not been established. Especially, the nature of the innermost region has not been clarified in a detailed and quantitative manner. 

\begin{figure*}[t]
\centering
\includegraphics[width=1.5\columnwidth]{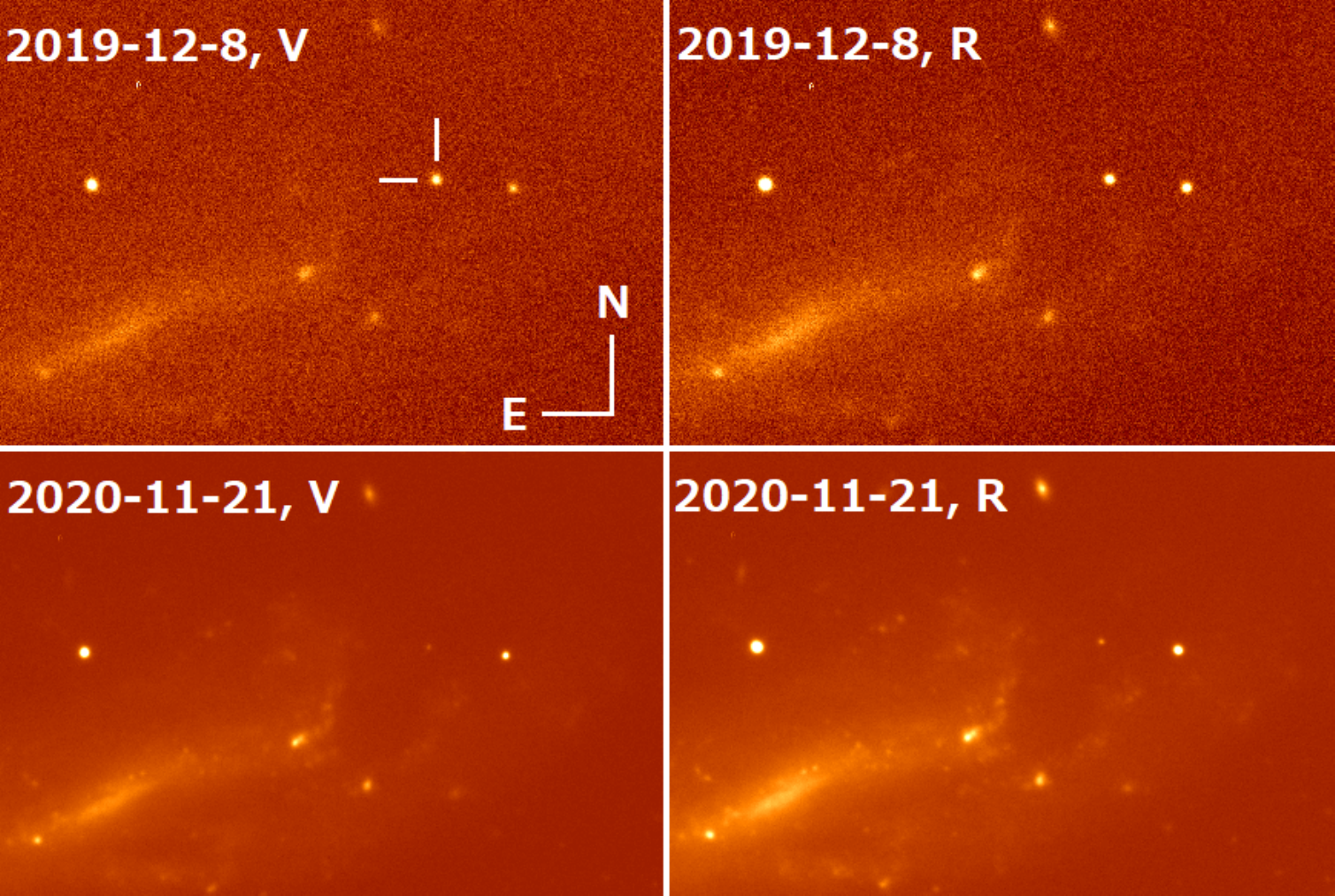}
\caption{The $V$ and $R$-band images of SN Iax 2019muj at 131 and 480 days after the explosion. 
}
\label{fig:images}
\end{figure*}

In the present work, we study properties of SN Iax 2019muj, which was discovered well before the maximum-light phase in a nearby galaxy \citep{barna2021,kawabata2021}. SN 2019muj has been classified as an intermediate class in its peak luminosity bridging between bright SN Iax objects (e.g., SNe 2002cx, 2005hk, 2012Z, and 2014dt) and subluminous ones (e.g., SNe 2008ha and 2010ae), with the early-phase spectral similarities to the subluminous SNe Iax. We especially focus on the late-phase behaviors, observational data of which are so far very limited especially for the fainter SNe Iax (including SN 2019muj). 

The paper is structured as follows. In Section 2, we present the late-time observations of SN 2019muj and data reduction. In Section 3, we overview characteristic spectral features of SN 2019muj, followed by the analyses of the line profiles of Ca lines ([Ca II]$\lambda\lambda$7292, 7324 and the Ca II NIR triplet) and allowed transitions (mainly Fe II), as well as possible detection of [O I]$\lambda\lambda$6300, 6363 (which can also be viewed as an upper limit). In Section 4, we provide comprehensive analyses of the spectral properties and their formation processes, including the allowed transitions (e.g., Fe II, Fe I, and the Ca II NIR) and the forbidden transitions (the [Ca II] and the [O I]), as supplemented by the spectral synthesis calculations for the allowed transitions. The analyses provide the first quantitative estimate on the nature of the innermost structure of SNe Iax, which provides a basis on discussing the origin of the inner dense core as presented in Section 5. We there propose a scenario to account for the late-time characteristic properties of SN 2019muj (and SNe Iax in general). The paper is closed in Section 6 with a summary of our findings.

\section{Observations and Data Reduction}\label{sec:obs}

\begin{figure*}[t]
\centering
\includegraphics[width=2\columnwidth]{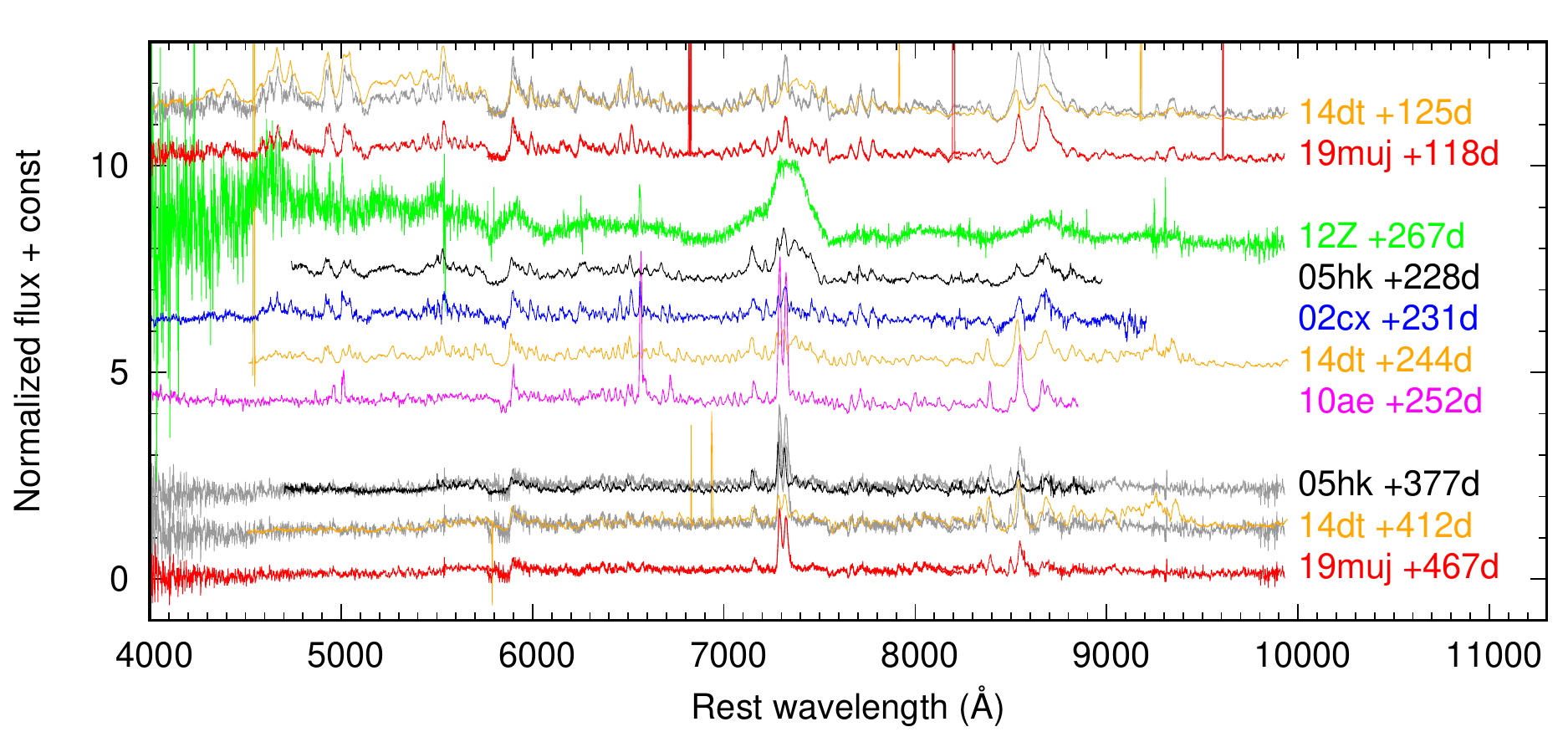}
\caption{The late-time spectra of SN 2019muj, as compared to other SNe Iax \citep{jha2006,sahu2008,stritzinger2014,yamanaka2015,kawabata2018}, on day $\sim +120$ days (top), $\sim +250$ days (middle), and $\sim +400$ days (bottom) since the maximum light. At a given phase, the spectra of different objects are shown roughly following an order of decreasing peak luminosity from top (bright) to bottom (faint). For spectra at $\sim +120$ and $\sim +400$ days, the corresponding spectra of SN 2019muj with an arbitrary flux scale are overplotted with each SN spectrum for comparison (gray). Note that the apparent H$_\alpha$ feature seen in a few SNe, especially strong in SN 2010ae, is likely a background host galaxy contamination.  The spectra for comparison SNe are obtained from WISeREP (https://www.wiserep.org) \citep{wiserep}.
}
\label{fig:spec_comp}
\end{figure*}

\begin{figure}[t]
\centering
\includegraphics[width=\columnwidth]{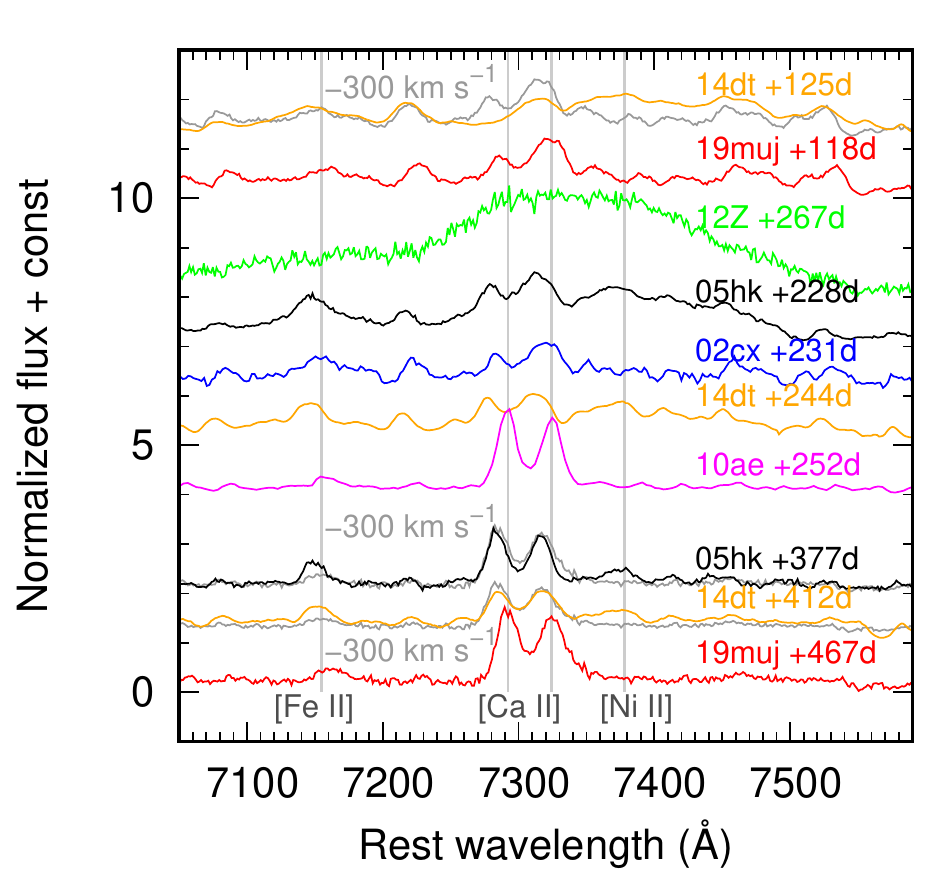}
\caption{The same as Fig. \ref{fig:spec_comp} but expanded around $7,300$\AA. The rest wavelengths of [Fe II]$\lambda$7155, [Ca II]$\lambda\lambda$7292,7324, and [Ni II]$\lambda$7378 are shown by the vertical gray lines. The spectra of SN 2019muj are overplotted by gray lines to the spectra of SN 2014dt (+125 and + 412 days) and SN 2005hk (+377 day), with an additional blueshift by $300$ km s$^{-1}$.  
}
\label{fig:spec_comp_ca}
\end{figure}

We have obtained late-time spectra of SN 2019muj with the 8.2m Subaru Telescope equipped with the Faint Object Camera and Spectrograph \citep[FOCAS;][]{kashikawa2002}, on 2019 December 8 (under the program S19B-055) and 2020 November 21 (S20B-056). The epochs correspond to +118 and +467 days after the maximum light, or 131 and 480 days after the explosion \citep[assuming the explosion date of 2019 July 30;][]{kawabata2021}. We note that the accurate estimate of the explosion date is not important for the purpose of the present work. The spectrum on day 131 has already been presented in \citet{kawabata2021}, while the one on day 480 is newly presented in this work. 

For both spectra, the same set up has been adopted. We used a 0.8" center slit, with the B300 (with no order-cut filter) and R300 (with the O58 filter) grisms. The set up covers the wavelength range of 3650–10000 Å. We used an atmospheric dispersion corrector with the slit direction set at P.A. $= 0$. The spectral resolution is $\sim 750$ as measured from the sky emission lines. The exposure times in each grism are 1,800 s (divided into two exposures per grism) on day 131, and 3,600 s (divided into four exposures per grism) on day 480, respectively. For flux calibration, Feige 34 (day 131) and Feige 110 (day 480) \citep{oke1990} were observed. 
\begin{figure*}[t]
\centering
\includegraphics[width=0.49\columnwidth]{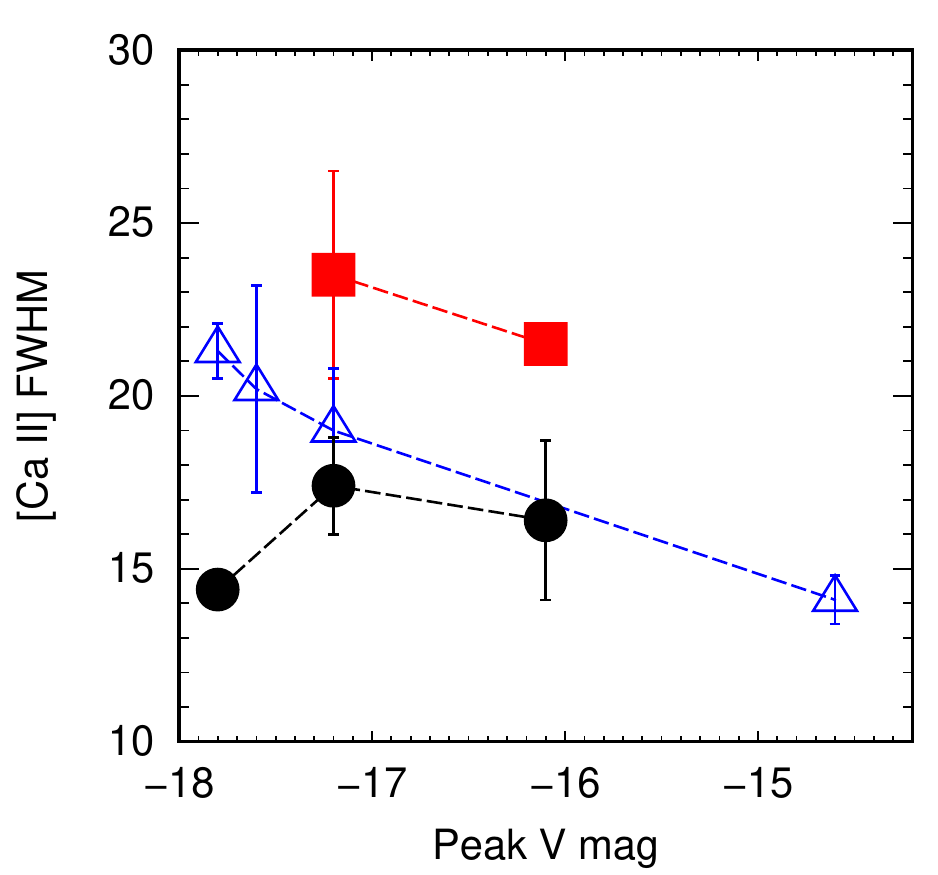}
\includegraphics[width=0.49\columnwidth]{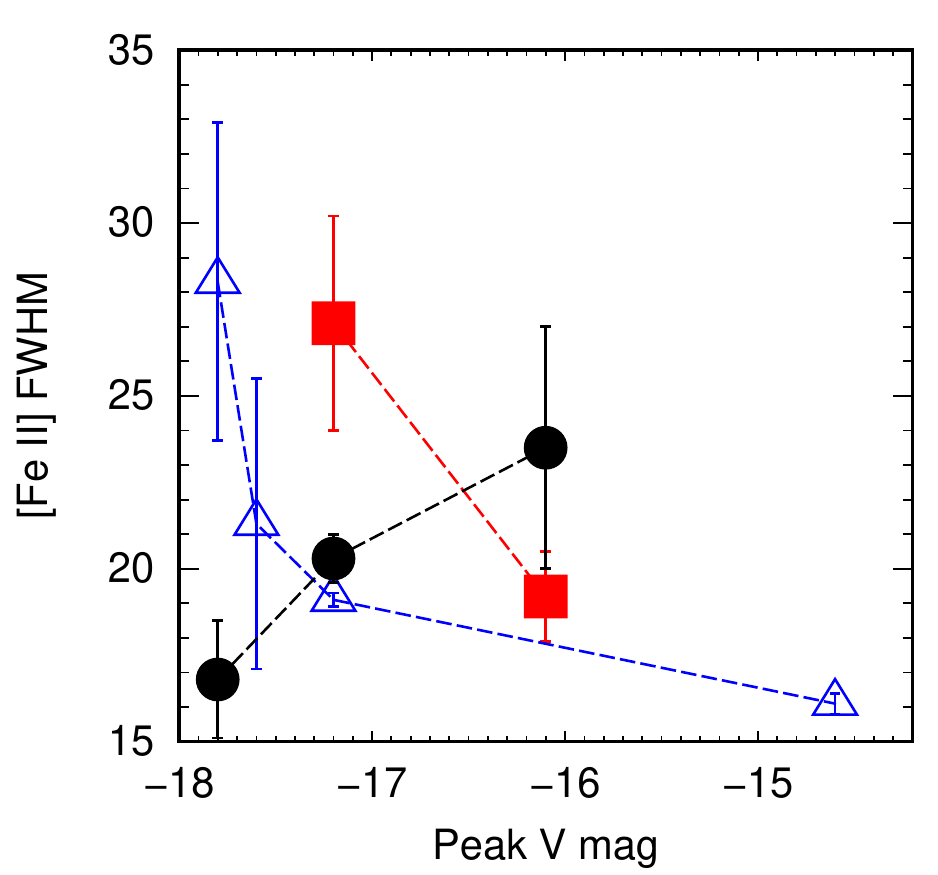}
\includegraphics[width=0.49\columnwidth]{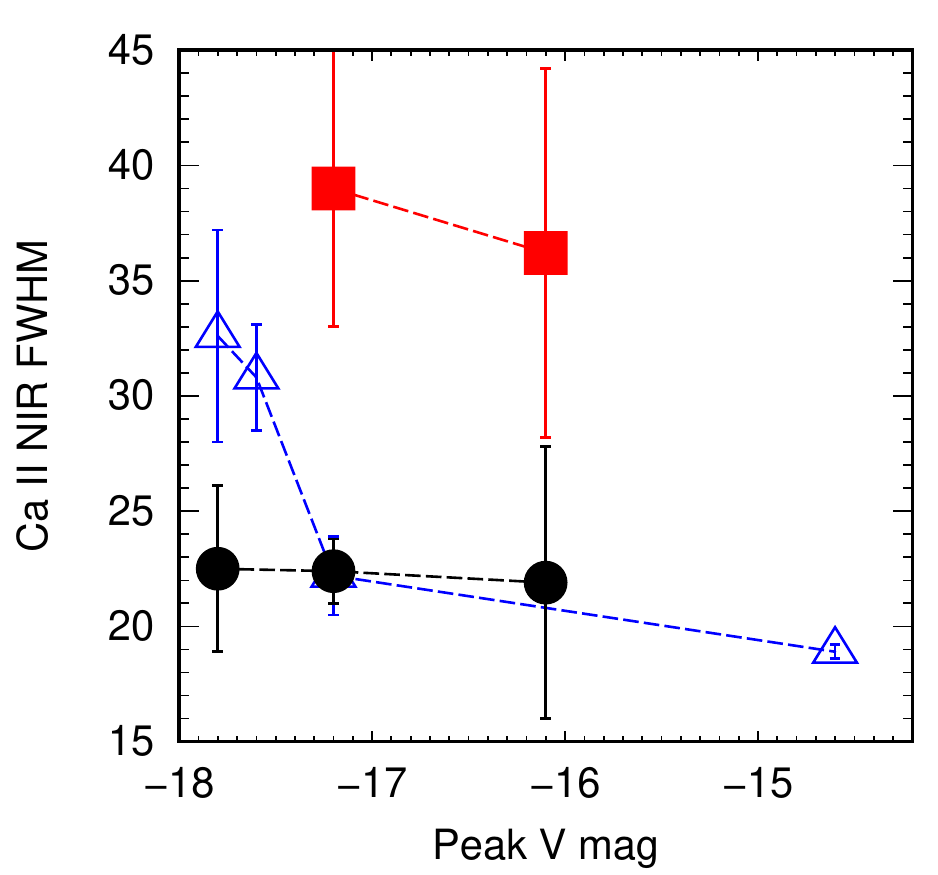}
\includegraphics[width=0.49\columnwidth]{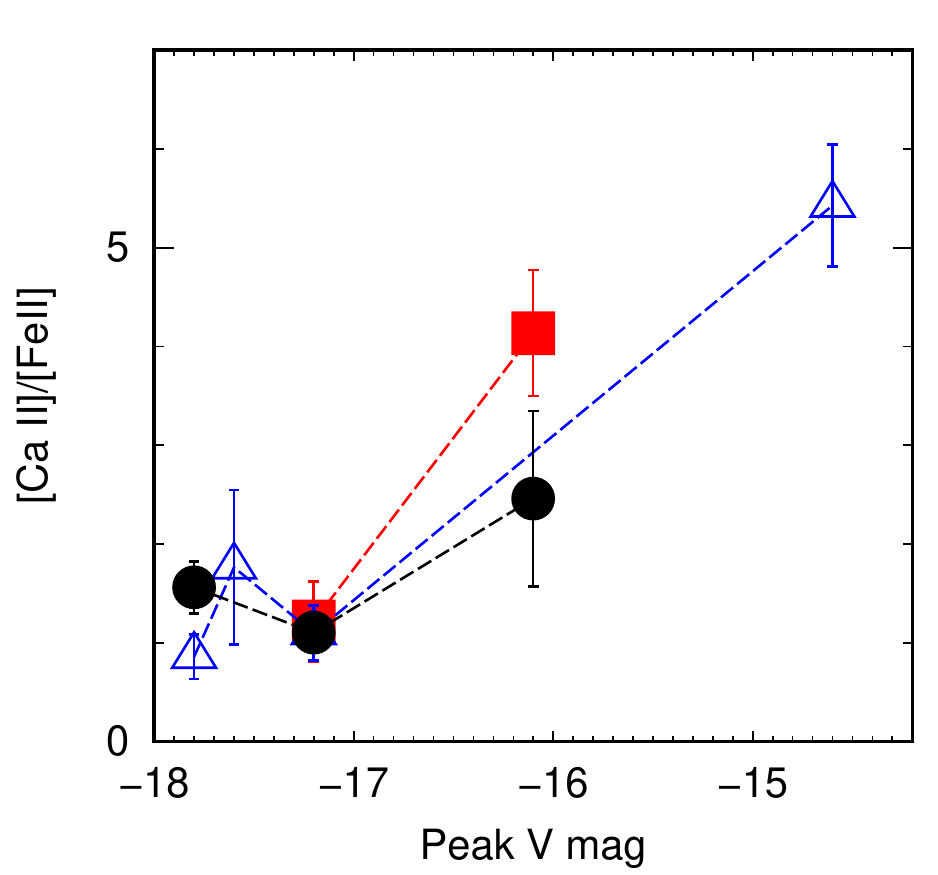}
\caption{Line widths (FWHM) of [Ca II]$\lambda$7324, [Fe II]$\lambda$7155, the 8542 component of Ca II NIR, and the flux ratio of [Ca II]$\lambda$7324/[Fe II]$\lambda$7155 (from left to right) for late-time spectra of the SN Iax sample shown in Figs. \ref{fig:spec_comp} and \ref{fig:spec_comp_ca} at $\sim +120$d (red; squares), $+250$d (blue; triangles), and $+400$d (black; circles). The dependence is shown as a function of the $V$-band peak magnitude, which is taken from the compilation by \citet{kawabata2021}.
}
\label{fig:neb}
\end{figure*}

The spectra were reduced following standard procedures with IRAF\footnote{IRAF is distributed by the National Optical Astronomy Observatory, which is operated by the Association of Universities for Research in Astronomy (AURA) under a cooperative agreement with the National Science Foundation.}, including bias subtraction, flat-fielding, cosmic-ray rejection \citep[with LAcosmic;][]{vandokkum2001}, sky
subtraction, 1D spectral extraction, wavelength calibration using ThAr lamp, and flux calibration \citep[e.g.,][]{maeda2015}. The wavelength solution was checked with the atmospheric sky
emission lines, and when necessary a small shift was further applied. Given that the late-time spectra of SN 2019muj show many narrow lines over the entire wavelength range, subtracting the telluric absorptions is not straightforward. The spectra shown in the present work are therefore not corrected for the telluric absorptions. 

The imaging observations were also conducted in the same nights with the Subaru/FOCAS, in the $V$ and $R$ bands. The exposure times are 20 sec in each band on day 131, and 180 sec in each band on day 480, supplemented with the shorter exposure images to avoid saturation of bright stars for relative photometry. The imaging data were reduced in a standard manner for
the CCD photometry, including the bias correction and flat-fielding. The relative photometry was performed in the same manner as described in \citet{kawabata2021}. The photometry results in $V=19.862 \pm 0.033$ and $R = 19.315 \pm 0.035$ on day 131 \citep[as already reported by][]{kawabata2021}, and $V=23.048 \pm 0.088$ and $R=22.095 \pm 0.044$ on day 480. Fig. \ref{fig:images} shows the Subaru/FOCAS images. 

The flux scales in the spectra were anchored with the results of the photometry. The flux was further corrected for the extinction within the Milky Way (MW). No correction was applied for the extinction within the host galaxy \citep{kawabata2021}. Finally, the flux is brought to the luminosity scale by applying the distance to the host galaxy, VV525, assuming the distance modulus of $\mu = 32.46 \pm 0.23$ mag \citep[see][]{kawabata2021}. The wavelength is corrected for the redshift of 0.007035 \citep{ann2015}. 

The spectra as compared to some SNe Iax in the late phases are shown in Figs. \ref{fig:spec_comp} and \ref{fig:spec_comp_ca}. In addition, in Appendix we show the spectra of individual frames together with the sky-emission and telluric-absorption patterns. 

\section{Late-time spectral evolution of SN 2019muj}

\subsection{Overall Properties}

Fig. \ref{fig:spec_comp} shows the comparison of the late-time spectra of SN 2019muj with a sample of SNe Iax at similar epochs. While we do not have a spectrum of SN 2019muj at $\sim +250$ days, a sample of SNe Iax in this phase is also shown. At each phase, the spectra are ordered roughly following the decreasing peak luminosity from top to bottom. Fig. \ref{fig:spec_comp_ca} shows an expanded view at $\sim 7,300$\AA. Fig. \ref{fig:neb} shows dependence of some characteristic features (Full-Width-Half-Maximum, FWHM, of [Ca II]$\lambda$7324, [Fe II]$\lambda$7155, and the 8542 component of Ca II NIR; the flux ratio of [Ca II]$\lambda$7324/[Fe II]$\lambda$7155) on the peak $V$-band magnitude.

As noted by some previous works, the late-time spectra are generally similar among the SN Iax class, despite the diverse natures in the early phase \citep{mccully2014b,stritzinger2014,foley2016}. We find that the similarity becomes more striking at more advanced epochs, which is seen by comparing the spectra of SNe 2019muj (intermediate/subluminous) and 2014dt (bright) at both phases ($\sim +120$ and $\sim +400$ days). At $\sim +120$ days, the spectral lines are broader for SN 2014dt \citep{kawabata2021}, following the early-phase behavior in which brighter objects show higher velocities. In the earlier phase ($\lsim +250$ days), a substantial difference is found at $\sim 7,300$\AA, which is the spectral region where SNe Iax show a substantial diversity \citep{stritzinger2014,stritzinger2015,foley2016},  and a clear trend is discerned in Figs. \ref{fig:spec_comp} and \ref{fig:spec_comp_ca} that this feature is related to the luminosity (see the feature at $\sim 7,300$\AA\ for the spectra at $\sim +250$ days ); a number of line features are clearly seen in the earlier epochs in the fainter SNe Iax (placed in the lower side in these figures), a convolution of which seems to create a bump-like broad feature seen in the brighter SNe Iax (placed in the upper side). On day $\sim +400$ days, all SNe Iax look nearly identical with the main difference in the strength of [Ca II]$\lambda7292,7324$ (see below for further details).

[Ca II]$\lambda\lambda$7292, 7324 is clearly present at $\sim +400$ days for all SNe Iax for which such late-time spectra are available. For most of the objects, the [Ca II] is also seen at $\sim 250$ days. The identification of  [Fe II]$\lambda$7,155 and [Ni II]$\lambda$7,378 is less certain. The features that might correspond to these forbidden lines are present in SNe 2005hk and 2014dt at $\sim +400$ days, with the velocity shift in the central wavelength roughly consistent with that seen in the [Ca II] (see below). The similar behaviors are seen at $\sim +250$ days except for SN 2012Z. We however note that numerous Fe II and Fe I transitions are present in this wavelength range, especially around the [Ni II] (Sections 4.2 and 4.3). As such, we believe that the [Fe II] identification is probably correct, but the [Ni II] identification may not be solid perhaps except for the most advanced phase.

The trends here, i.e., the correlation between the late-time spectral features and the luminosity, and the convergence of the spectra toward the later epochs, are also seen in Fig. \ref{fig:neb}. The forbidden lines are generally narrower for fainter SNe Iax until $\sim +250$ days, while they are similar in the latest epochs ($\sim +400$ days) or even show a hint of the inverted trend. The width of the Ca II NIR is substantially larger than those of the forbidden lines at $\sim +120$ days. The Ca II width shows quick decrease by $\sim +250$ days in the fainter objects, followed by a similar decrease toward $\sim +400$ days in the brighter objects, finally reaching to similar widths for all SNe Iax; the epochs to reach to the `converged' spectra are thus dependent on the luminosity. Finally, the [Ca II]/[Fe II] ratio does not evolve much as a function of time; it indicates that the evolution of the SN Iax late-time spectra is mainly attributed to the decreasing contribution from the region emitting the allowed transitions. For the converged `nebular' spectra, a clear trend is seen in the [Ca II]/[Fe II] ratio as a function of the luminosity; [Ca II] is stronger relative to [Fe II] in the fainter objects. 

The comparison here shows the following properties. (1) The late-time spectral diversity seems to from a sequence following the peak luminosities. (2) The difference is becoming smaller at more advanced epochs, and converges to very similar spectra. (3) The transition to this `generic' (converged) late-time spectrum takes place earlier for the fainter objects. (4) The main difference seen in the `converged' spectra is the strengths of the [Ca II] and [Fe II]+[Ni II], the property of which seems to follow the luminosity sequence in a way that the stronger [Ca II] and weaker [Fe II]+[Ni ii] are associated with fainter SNe Iax. 

The behaviors (3) and (4) are especially evident in the spectral evolution of SNe 2010ae and 2019muj. The behavior (4) as found here might have an interesting implication for the explosion mechanism; [Fe II]$\lambda7155$ and [Ni II]$\lambda7378$ have been suggested to arise in the high-density ash of the deflagration flame for normal SNe Ia \citep{maeda2010b,maeda2010c}, and the same argument may also apply to the failed/weak deflagration model for SNe Iax. This interpretation then suggests that the brighter SNe Iax, with the stronger [Fe II] and [Ni II] in the late phase, may have a larger amount of the high-density deflagration ash in the innermost region. 

Another interesting property is the velocity shift. While the sample is small, SNe Iax tend to show the blueshift in the [Ca II] (and also in the [Fe II]) with no clear examples (at least in the sample presented here) showing the redshift. As time goes by, the velocity tends to decrease (i.e., moving redward) \citep[see also][]{foley2016}. We note that a part of the line shift in the earlier phase may be attributed to absorption by allowed transitions, as demonstrated for the [Ca II] in case of SN 2019muj in the present work (Section 3.2).

The shift up to $\sim 300$ km s$^{-1}$ is still visible at $\sim +400$ days, which may trace the intrinsic kinematical behavior (see Section 3.2). Only four SNe Iax, SNe 2010ae at $\sim +250$ days and SNe 2005hk, 2014dt, and 2019muj at $\sim +400$ days, seem to enter into such a regime (Fig. \ref{fig:spec_comp_ca}), and thus it is not yet clear whether the apparent imbalance in the shift (no shift for two SNe and blueshift of $\sim 300$ km s$^{-1}$ for the other two, with no example for redshift) is real or just due to statistical fluctuation.  It is important to add further samples in such sufficiently late phases to search for SNe Iax with redshifted forbidden lines. In any case, the finding adds another item in the list of the observed behaviors; (5) the fainter SNe Iax tend not to show the shift in the central wavelength in the forbidden lines. Together with the items (3) and (4) above, the properties of the forbidden lines seem to form a sequence as correlated with the peak luminosity.

If the behavior (5) might be kinematical in its origin (which need to be tested by future observations), it should provide another interesting implication for the progenitor systems or explosion mechanisms. While this is the level of the global motion of the region including the progenitor systems (e.g., the host rotation), it might also reflect either an asymmetry in the explosion or orbital motion of a possible binary progenitor. The item (5) above, i.e., the faintest SNe Iax do not show the shift in the central wavelength, is qualitatively the opposite to the expected asymmetry in the ejecta for the failed/weak deflagration scenario, where a weaker explosion is likely associated with stronger asymmetry created by the initial deflagration \citep{maeda2010,fink2014}. The orbital motion of a binary progenitor, on the other hand, could provide a natural interpretation. If we would assume a binary of a $M_{\rm Ch}$ WD and a Roche-lobe filling He star with a comparable mass \citep[i.e., the system suggested for a possible companion detection in SN 2012Z;][]{mccully2014a}, the orbital motion is $\sim 300$ km s$^{-1}$. Then, if the shift would be attributed to the progenitor orbital motion, it would mean that the fainter SNe Iax may be associate with binaries with wider separation, which may be related to a lower accretion rate to the primary. 

\begin{figure}[t]
\centering
\includegraphics[width=\columnwidth]{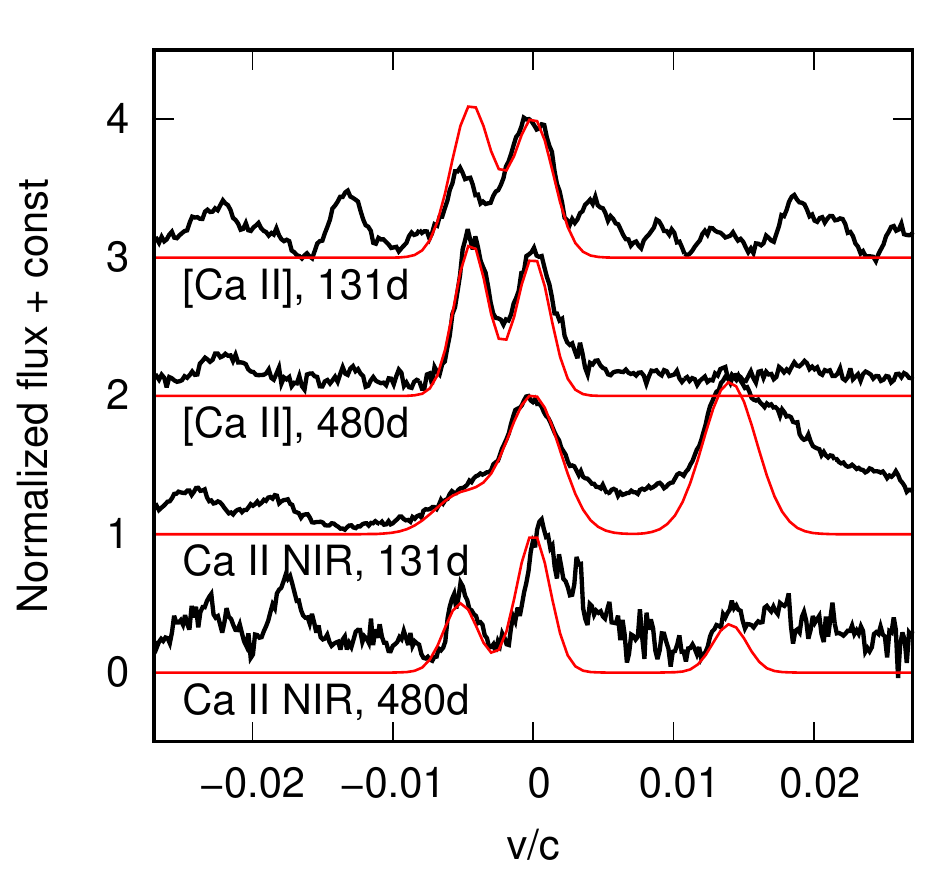}
\caption{Line profiles of [Ca II]$\lambda\lambda$7292, 7324 and the Ca II NIR triplet, on day 131 and day 480 (black). The convolution of Gaussian profiles are shown (red). The Gaussian FWHM is taken to be the same for the $\lambda7292$ and $7324$ components at a given epoch; 900 km s$^{-1}$ (day 131) and 760 km s$^{-1}$ (day 480), after being corrected for the instrumental spectral resolution. The same is the case for the $8498$, $8542$, and $8662$ components of the Ca II NIR triplet; 1,300 km s$^{-1}$ on day 131 and 760 km s$^{-1}$ on day 480. 
}
\label{fig:ca_profile}
\end{figure}

As investigated here in this section, late-time spectra provide various insights into the origin(s) of SNe Iax. Similarities and diversities, together with possible relations between the early-phase and late-phase properties, are found. These properties should ultimately be explained in a unified manner by a single scenario (unless SNe Iax are a complete mixture of totally different populations), but one of course has to solve the problems one by one. In the rest of the present work, we mainly focus on the question related to the similarity of SNe Iax in the late phase; by studying the details of the late-time spectra (and light curve) of SN 2019muj, we try to answer how the main features of late-time spectra of SNe Iax as distinct from normal SNe Ia are explained.

\subsection{[Ca II] and Ca II NIR lines}

Fig. \ref{fig:ca_profile} shows an expanded view around [Ca II]$\lambda\lambda7292$, $7324$ and the Ca II NIR triplet in the velocity space. The figure also shows a convolution of Gaussian profiles, in which the central wavelengths are set at the rest wavelengths of the lines for all the components. The FWHM of the input Gaussian profile is set equal for the two components of the [Ca II] and for the three components of the Ca II NIR. We however allow different values of the FWHM for the [Ca II] and the Ca II, and also at different epochs. 

The forbidden lines should trace the intrinsic kinematical property by the line profile better than the allowed lines. Also, possible radiation transfer effects should become smaller for a more advanced epoch. We therefore first examine the [Ca II] profile on day 480. We find that the line centers are consistent with having no shift with respect to the rest wavelengths. The FWHM of 850 km s$^{-1}$ can reproduce the observed line profile reasonably well. The FWHM after being corrected for the instrumental spectral resolution corresponds to 760 km s$^{-1}$. 

Applying the same procedure for the [Ca II] profile on day 131, we find that different amounts of the line-center shift are required for the 7292 and 7324 components. This is demonstrated in Fig. \ref{fig:ca_profile}, where the convolution of the Gaussian profiles, centered on the rest wavelengths of these components, is compared with the [Ca II] profile. The red component (7324\AA) indeed does not change in its central wavelength as compared to day 480 (i.e., no shift), while the blue one (7292\AA) is shifted to the blue as compared to the profile on day 480. In addition, the relative strength of the 7292 component to the 7324 component is weaker on day 131 than on day 480. The difference in the central wavelengths should not reflect intrinsic difference in the kinematical property of the line-emitting region, as the two components are both coming from essentially the same transition, i.e., 3d$^{2}$D - 4s$^2$S. Further, the relative strength of the two components should not evolve much for the same reason. 

Given that the 7292 component shows both the shift in the wavelength and the suppression in the flux on day 131 as compared to day 480, the most likely interpretation is that the 7292 component suffers from an absorption by some allowed transitions. We identify Fe II 7310, Fe II 7312, and Fe II 7323 as candidates for such allowed transitions, which are confirmed to dominate this wavelength region with comparable strengths, later through the analysis of the physical conditions for the allowed transitions (Sections 4.2 and 4.3). We measure that the `absorption minimum' is $\sim 7299$\AA, which is $\sim 650$ km s$^{-1}$ blueward of the average wavelength for the three Fe II transitions (7315\AA). The velocity is smaller than (but roughly similar to) the width of the [Ca II] on the same epoch. Therefore, if the allowed transitions are formed in a similar kinematical structure with the forbidden line-forming region, the peculiar wavelength shift in the 7292 component on day 131 is explained by the absorption by the Fe II transitions, whose effect decreases to become negligible on day 480. Inversely, this finding suggests that the forbidden lines ([Ca II]) and the allowed lines (Fe II) are formed largely in the same region, with the allowed line-forming region indeed more extended to the outer region than the forbidden line-forming region on day 131 (see below). This is against a naive expectation attributing the forbidden line-forming region to an outer, less dense region. 

The line width (intrinsic FWHM) of the [Ca II] as estimated from the Gaussian profiles evolves from 900 km s$^{-1}$ on day 131 to 760 km s$^{-1}$ on day 480. A similar evolution is seen for the Ca II NIR triplet. The profiles of all the components (8498, 8542, and 8662\AA) are consistent with no shift in the central wavelengths, the property that is shared with the [Ca II], in both epochs. The Gaussian profile shown in Fig. \ref{fig:ca_profile}, which provides a reasonable representation of the observed profiles (given contamination from other allowed transitions, e.g., Fe II), has the intrinsic FWHM evolving from 1,300 km s$^{-1}$ on day 131 to $760$ km s$^{-1}$ on day 480. On day 131, the Ca II NIR has a broader line width than the [Ca II], indicating that a relatively low-density outer region was initially opaque to the Ca II NIR but became transparent on day 480. The line widths converge to the same value on day 480 for the [Ca II] and the Ca II NIR, suggesting that an inner component, which has the characteristic velocity of $\sim 760$ km s$^{-1}$ and dominates the spectral formation, is exposed on day 480. This behavior further suggests that there is a jump in the density structure at $\sim 700-900$ km s$^{-1}$, i.e., the inner component is `distinct' from the outer ejecta; otherwise, we expect to see a continuous change in the line widths in which the Ca II NIR keeps broader than the [Ca II].

\begin{figure}[t]
\centering
\includegraphics[width=\columnwidth]{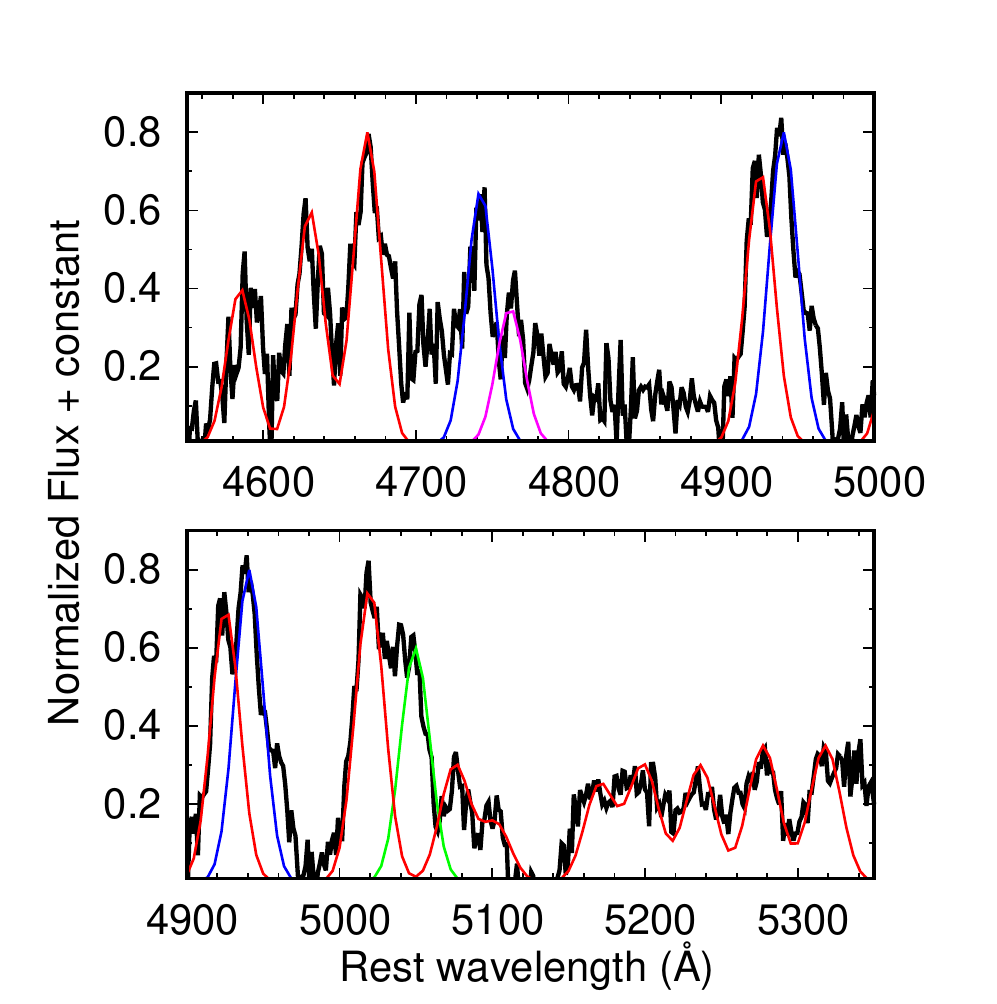}
\caption{The spectrum of SN 2019muj on day 131 in 4550-5350\AA. Plotted here for comparison are the same Gaussian profiles for some strong allowed Fe II (red), Co II (blue), Mg II (magenta), and Si II (green) lines, adopting the identical profile with that for the Ca II NIR on day 131.
}
\label{fig:fe_profile}
\end{figure}

\subsection{Fe II and allowed transitions}

Figure \ref{fig:fe_profile} is an expanded view (4550-5350\AA) of the spectrum on day 131. This wavelength range shows many strong features, and thus it is shown as an example of line identifications and further consideration on the kinematical structure. In Fig. \ref{fig:fe_profile}, we have selected some strong allowed transitions (see Sections 4.2 and 4.3), and compute the convolution of the line profiles, each of which is assumed to have the same Gaussian profile as the one derived for the Ca II NIR (Fig. \ref{fig:ca_profile}) (with the line strengths set arbitrarily). The Gaussian line-profile model is shown for the following transitions; Fe II 4585.1, 4630.6, 4668.1, 4925.3, 5019.8, 5075.5, 5102.1, 5170.4, 5199.0, 5236.1, 5277.4, and 5318.1; Co II 4742.3 and 4940.7;  Mg II 4757.1 \& 4766.1 (with the central wavelength taken as 4761.6\AA); Si II 5042.4 \& 5057.4 (with the central wavelength taken as 5049.5\AA). We find that the spectrum is dominated by Fe II \citep[see also][]{branch2004,jha2006,sahu2008,mccully2014b}, with other contributions provided by Co II, Mg II, and Si II in this wavelength range. Also, it is seen that the simple Gaussian profile model reproduces the observed line profiles reasonably well. From this analysis, we conclude that the strong spectral features are dominated by the allowed transitions mainly from Fe peak elements with the kinematical structure identical to that derived by the Ca II NIR. 

We note that this exercise here is done for a demonstration purpose. In practice, many weak transitions can make major contributions in the expanding ejecta \citep[i.e., the expansion opacity;][]{karp1977,eastman1993}. In addition, the emerging spectral feature is a convolution of the emission and absorption features, and thus assuming the same line profile for all the lines is a rough approximation. We will take into account these effects in the subsequent analyses (Section 4.2 for the line identification and Section 4.3 for the spectral formation).

\subsection{[O I]}

While there has been possible identifications of allowed O I transitions in late-time spectra of SNe Iax \citep[e.g.,][]{jha2006}, there has been no report on detection of [O I]$\lambda\lambda$6300, 6363 in late-phase spectra of SNe Iax to our knowledge. Indeed, the non-detection of the [O I] has been regarded to be a challenge for models including a large amount of oxygen in the ejecta, such as a weak but successful WD thermonuclear explosion scenario \citep[e.g., energetic models in][which leave no bound remnant]{fink2014} or a core-collapse SN scenario. This may possibly be remedied if oxygen-rich materials are microscopically mixed with iron-rich or Si-rich materials \citep{maeda2007,sahu2008}, but such a mixing process is unlikely to be realized during the explosion. 

\begin{figure}[t]
\centering
\includegraphics[width=\columnwidth]{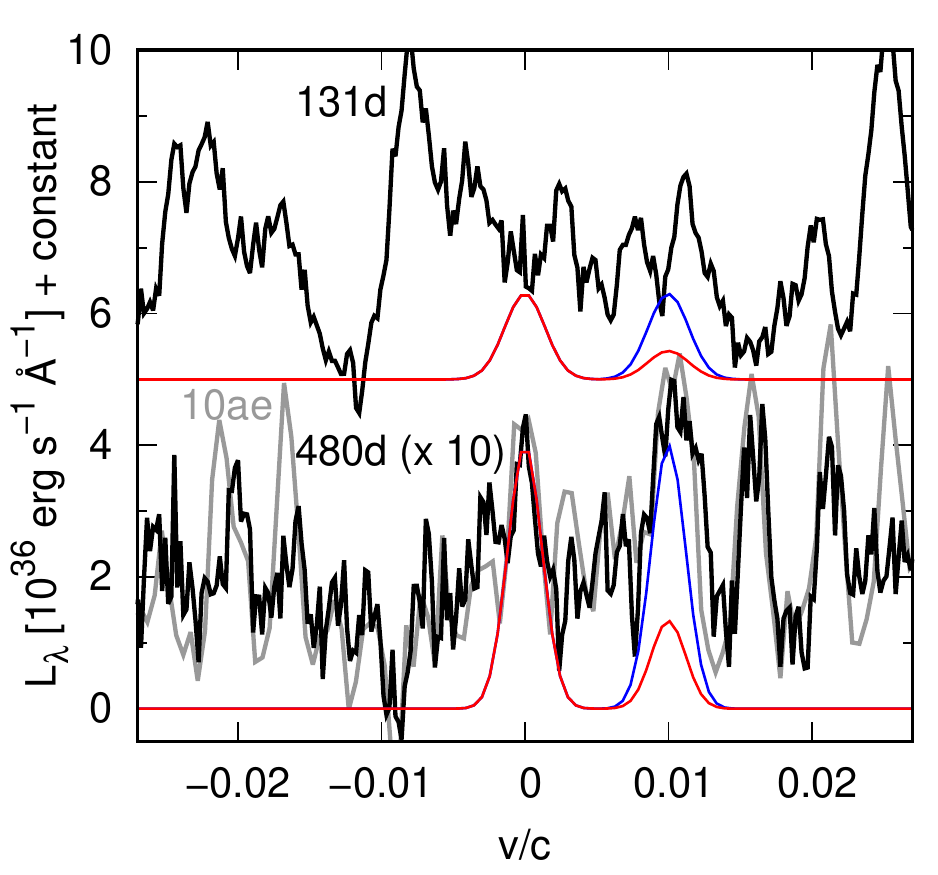}
\caption{The spectra around [O I]$\lambda\lambda$6300, 6363, on day 131 and day 480 (black). To place upper limits on the fluxes, we plot the Gaussian profiles adopting the same line profile with the [Ca II] but with the flux set not to exceed the observed flux. The line ratio between the two components is fixed to be 3:1 (red) or 1:1 (blue). The flux scale for the spectrum on day 480 is multiplied by a factor of 10 for a visualization purpose. Also plotted for comparison is a spectrum of SN Iax 2010ae at $+252$ days since the maximum light (gray) with an arbitrarily-scaled flux level. 
}
\label{fig:o_profile}
\end{figure}

We find that a close look at the spectra of SNe Iax indeed shows possible detection of the [O I]. Fig. \ref{fig:o_profile} shows the constraint on [O I]$\lambda\lambda$ 6300, 6363 for SN 2019muj. The same Gaussian line-profile model with that used for the [Ca II] is plotted, with the line flux set so as not to exceed the observed flux. On day 131, the [O I]$\lambda\lambda6300$, $6363$ doublet is not detected; the upper limits for the fluxes are derived to be $\sim 3.0$ and $\sim 1.0 \times 10^{37}$ erg s$^{-1}$ for the 6300 and 6363 components, respectively. On day 480, there are features which match to [O I]$\lambda\lambda6300$, 6300 in their wavelengths; from the Gaussian line-profile model, we derive the fluxes of these features to be $\sim 7.6 \times 10^{36}$ erg s$^{-1}$ for both components. We note that similar features are present also in SNe 2005hk and 2014dt at a similar epoch ($\sim +400$ days) and also in SN 2010ae at an earlier epoch ($+252$ days) (see Fig. \ref{fig:o_profile}), following the striking similarity of the SN Iax spectra toward the later phase (Section 3.1). As such, the analysis in the present work (see Section 4.5 for details) is indeed applicable to other SNe Iax as well. 

We however regard these features on day 480 just as possible [O I] detection. Given that there are many spectral features seen in the spectra, the identity of the features at $\sim 6300$\AA\ and $6363$\AA\ is not very clear. Conservatively, the derived fluxes on day 480 provide upper limits on the strengths of the two components of [O I]$\lambda\lambda$ 6300, 6363. 

\section{The physical properties of the inner region}

In this section, we derive the physical properties of the inner ejecta as probed by the late-time spectra. For SN 2019muj, there is an outer ejecta model as constrained with the early-phase spectra. First, we perform spectral synthesis for an ejecta model that is extrapolated inward from the early-phase model, showing that the model does not reproduce the late-time spectra (Section 4.1). Then, we consider the optical depths of Fe II and I transitions, and conclude that there must be a high-density inner component which is distinct from the outer ejecta, to explain a numerous number of spectral features seen in the late-time spectra. We also constrain the mass of Fe (which can be initially $^{56}$Ni) in the inner component (Section 4.2). The idea is further tested with the spectral synthesis simulations, confirming that such a model can reproduce major spectral features (Section 4.3). The investigation up to this point does not constrain much on the compositions except for Fe and Fe-peak elements. We then turn into the analyses of forbidden lines, [Ca II] in Section 4.4 and [O I] in Section 4.5; we show that the Ca and unburnt oxygen are only subdominant, e.g., the mass fraction of oxygen is constrained to be $\lsim 5$\% of the total mass of the inner component. These analyses thus strongly constrain the physical properties of the inner ejecta. Based on the findings, the origin of this inner component and the explosion mechanism of SNe Iax will be discussed in Section 5. 

\subsection{A few critical issues}

\begin{figure*}[t]
\centering
\includegraphics[width=1.0\columnwidth]{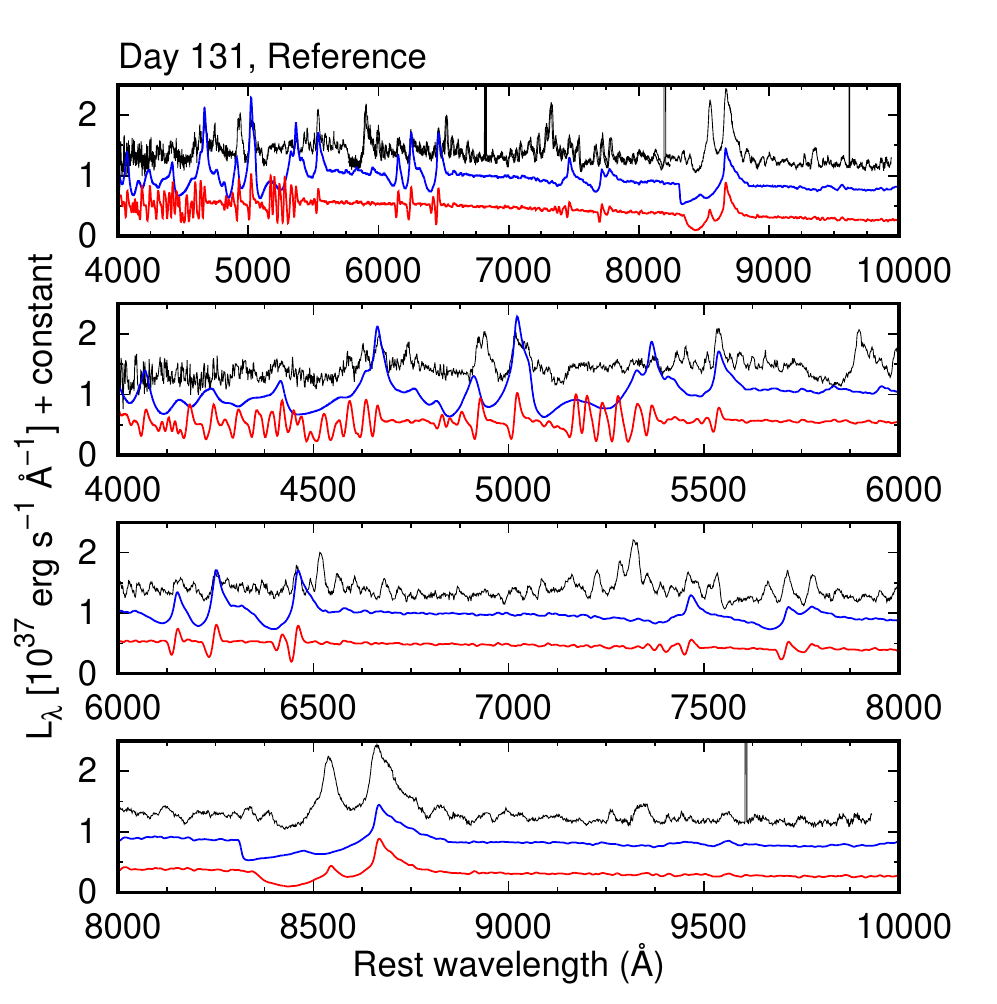}
\includegraphics[width=1.05\columnwidth]{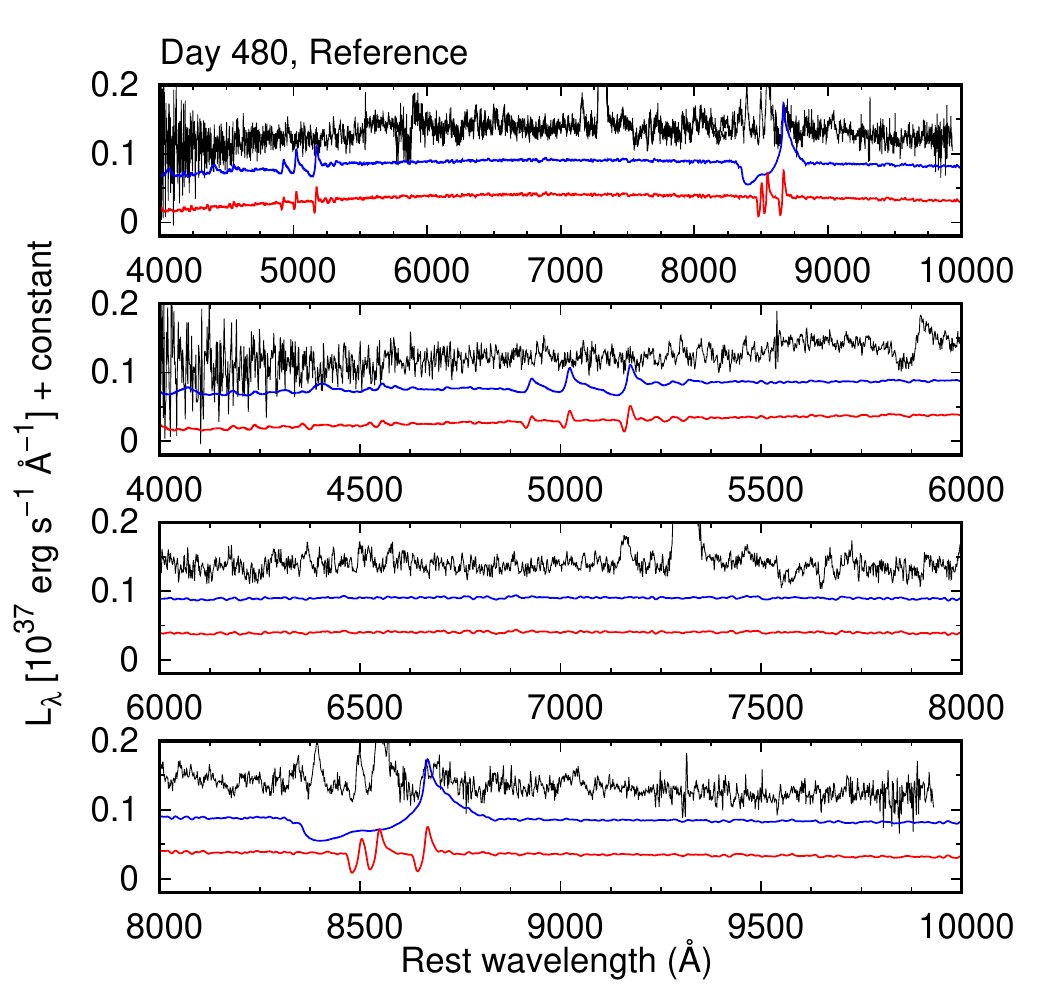}
\caption{The TARDIS spectral-synthesis models on days 131 (left) and 480 (right), for the reference density/abundance structure extrapolated from the analysis for the earlier epochs (blue). Also shown is the same model but with the abundance suppression of Ca and heavier elements above $1,000$ km s$^{-1}$ (red). As the model input, the bolometric luminosity and the photospheric velocity are set as follows; $10^{7.05} L_\odot$ and $230$ km s$^{-1}$ on day 131; $10^{6.05} L_\odot$ and $35$ km s$^{-1}$ on day 480. The photospheric temperature is $\sim 5500$ K on day 131 and $\sim 4100$ K on day 480. }
\label{fig:model_ref}
\end{figure*}

To set a scene for further analyses of the late-time spectra of SN 2019muj, we first address the predictions from the model constructed based on the early-phase spectra. \citet{barna2021} performed such an analysis, through which they constructed an ejecta model with an exponential density profile (truncated at a high velocity). Specifically, their (homologously expanding) ejecta model is described by the central density of 0.55 g cm$^{-3}$ $(t/100 \ {\rm s})^{-3}$ and the e-folding velocity of 1,800 km s$^{-1}$ (with the truncated velocity of 5,500 km s$^{-1}$). The central density corresponds to $3.8 \times 10^{-16}$ g cm$^{-3}$ on day 131 and $7.7 \times 10^{-18}$ g cm$^{-3}$ on day 480, respectively. This density structure is consistent with the expectation for the unbound ejecta in the failed deflagration scenario; it is between the model N1def of \citet{fink2014} and the model N5def\_hybrid of \citet{kromer2015}. 

We perform spectral synthesis calculations adopting the same density structure. For the composition, we assume a uniform abundance structure, adopting the model N1def of \citet{fink2014} for the mass of each element in the unbound ejecta; $\sim 40$\% in the mass fraction is $^{56}$Ni, $\sim 30$\% is the C+O composition, $\sim 10$\% is for a combination of stable Fe and Ni, $\sim 8$\% is for a combination of Si and S, and the remaining fraction is distributed to Mg, Ca ($\sim 0.3$\%) and other Fe-peak elements. Hereafter, we call this ejecta model the reference model. Note that the model has not indeed been strongly constrained below $\sim 3,500$ km s$^{-1}$, which corresponds to the photospheric velocity used to model the spectrum at the last epoch  studied by \citet{barna2021}. 

We use the TARDIS \citep{kerzendorf2014} for the spectral synthesis calculations in the present work, which assumes a sharp photosphere with its velocity (i.e., the radius in the homologously expanding ejecta) and temperature as input parameters. The options for the physics modules are set as follows; `nebular' for ionization, `dilute-lte' for excitation, `detailed' for radiative rate, and `macroatom' for line interaction. The number of the photon packets is set as $10^5$ during the iterations while it is set as $10^6$ packets (each divided further to 10 virtual packets) in computing the final spectrum for each model. We further convolve a Gaussian profile to the synthesized spectrum with the FWHM corresponding to the instrumental wavelength resolution, since the lines are narrow and the instrumental broadening must be taken into account. 

Fig. \ref{fig:model_ref} shows the synthesized spectra for the reference model on day 131 and day 480. While the model (blue) shows some strong lines (especially on day 131) as observed, the similarity of the synthesized spectra to the observed spectra is very poor in both epochs; (1) the model predicts too smooth spectra without many lines that are however clearly detected in the observational data, and (2) the strong lines seen in the models are associated with blue-shifted absorption toward the high velocity that is inconsistent with the data. The point (2) can for example be seen in the Ca II NIR triplet, where the 8542 component in the model is totally suppressed by the blue-shifted absorption tail of the 8662 component. 

\begin{figure}[t]
\centering
\includegraphics[width=1.0\columnwidth]{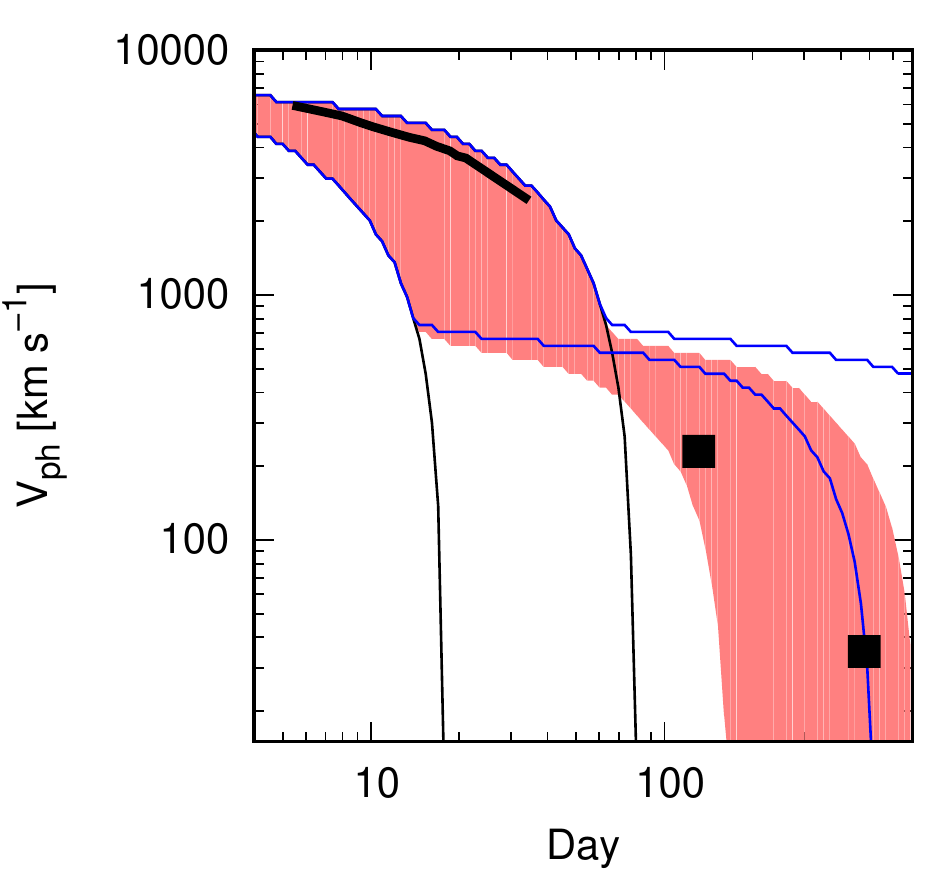}
\caption{A roughly expected range of the photospheric velocity as a function of time. For the reference model (thin black; Section 4.1) and the C+O-rich inner component model (blue; see Section 4.3), two curves are shown for each model with a constant opacity of 0.025 g cm$^{-1}$ and 0.5 g cm$^{-3}$. For the $^{56}$Ni-rich model (Section 4.3), the same range is shown by a red-shaded area. For comparison, the photospheric velocity evolution derived by \citet{barna2021} for the early phase is shown by a thick black curve, and the photospheric velocities in the late phase adopted in the TARDIS calculations in the present work (Section 4.3) are shown by black-filled squares. 
}
\label{fig:photosphere}
\end{figure}

The problem (2) can simply be an artifact in the simulation. The model density in the outer region becomes low in the late phases analyzed here, and the treatment of the ionization and excitation is not well justified. This can affect the strength of the lines; for example, the strength of the Ca II NIR can be very sensitive to the treatment of line transfer and the non-Local Thermodynamic Equilibrium (LTE) effect due to its low ionization potential and large oscillator strength \citep{kasen2006}. Given that the reference model is indeed not reliable below $\sim 3,500$ km s$^{-1}$ (which we want to constrain by the late-phase spectra) and we see the typical velocity of 760 - 1,300 km s$^{-1}$ in the observed spectra (Sections 3.2 and 3.3), we phenomenologically modify the reference model, by suppressing the mass fractions of Ca and heavier elements above 1,000 km s$^{-1}$ by a factor of 100 (which is compensated by increasing the mass fractions of C and O equally). This model (red lines in Fig. \ref{fig:model_ref}) shows some improvement; the lines in the synthesized spectra are now as narrow as observed, with the high-velocity absorption removed. Also, the blue portion of the model spectrum on day 131 now shows many narrow lines with some similarity to the observed features. However, the synthesized spectrum on day 131 shows little line features above $\sim 5,500$\AA\ except for the Ca II NIR. On day 480, the model spectrum does not show much improvement (except for the Ca II NIR), with little line features predicted in the model. 

To evaluate the uncertainty, we further test a range of the photospheric luminosity and velocity. The photospheric luminosity is varied by $\sim \pm 30$\%, and the photospheric velocity is so by $\sim \pm 15$\% (on day 131) and $\sim \pm 40$\% (on day 480), beyond which the flux level and the continuum color substantially deviate from what are observed. We see no improvement in the spectral features. 

\begin{figure*}[t]
\centering
\includegraphics[width=1.05\columnwidth]{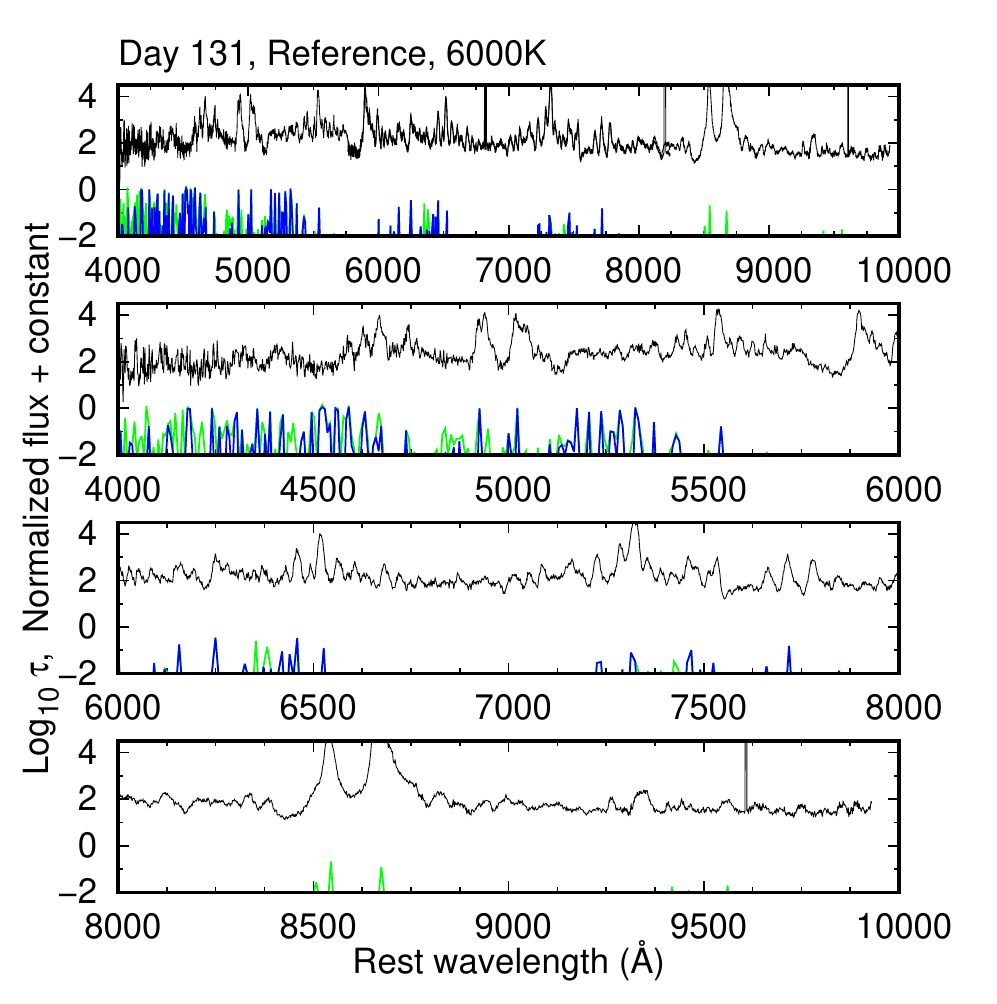}
\includegraphics[width=1.05\columnwidth]{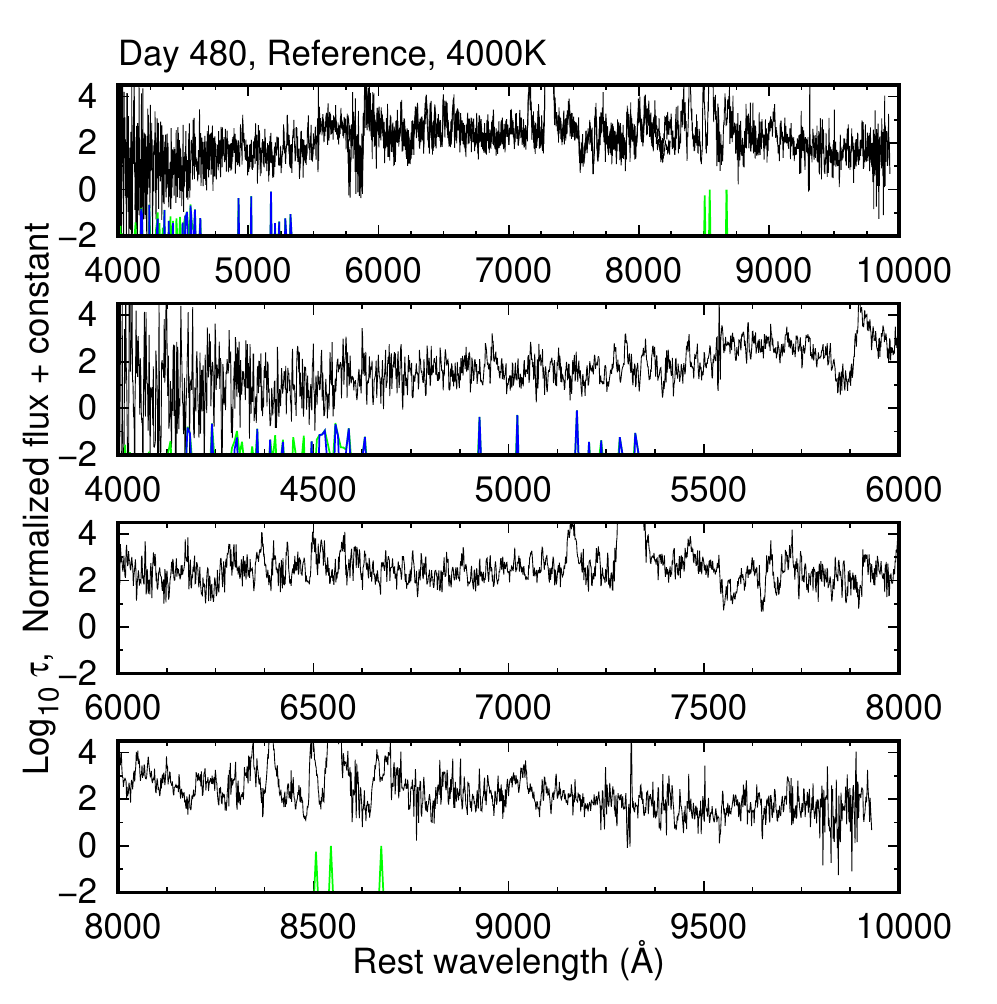}
\caption{The distribution of the optical depths of the allowed transitions based on the expansion opacity formalism, for the reference model on days 131 and 480. The contributions from Fe II (blue) and Fe I (red; rarely seen in this model) are shown, as well as the total optical depth including contributions from other species (green). For comparison, the observed spectra of SN 2019muj at the corresponding epochs are shown with an arbitrary flux scale. 
}
\label{fig:onezone_ref}
\end{figure*}

Indeed, due to the low density in the inner region of the reference model, the formation of the photosphere also has a problem. Fig. \ref{fig:photosphere} shows the expected range of the photospheric velocity as a function of time for the reference model, together with two additional models which are to be investigated later (Section 4.3). A constant opacity is assumed here; the typical opacity for SNe Iax is uncertain, and we conservatively consider a range of the opacity between 0.025 and 0.5 g cm$^{-3}$ \citep[e.g.,][]{mazzali2001,maeda2009a} covering the Si/S-rich and Fe/Ni-rich compositions. It is seen that the photosphere expected in the reference model recedes very quickly once it drops below $\sim 1,000$ or $2,000$ km s$^{-1}$, i.e., the characteristic velocity in the exponential density distribution in the reference model. The photospheric velocity in the late phases should be much lower than this value as is clear from the line-profile analysis. It is very difficult to keep the photosphere at such a low velocity for nearly 500 days. Note that this argument is not sensitively dependent on the exact position of the photosphere. In reality, the opacity is also dependent on the ionization and thermal conditions. However, even taking this uncertainty into account, the rapid recession in the photosphere is generally expected for the nearly constant inner density found for the reference model. Therefore, the reference model as it is does not explain the characteristic behavior of the late-time spectral evolution of SN 2019muj (and SNe Iax in general), i.e., the long-lasting appearance of the photosphere and associated spectral features dominated by allowed transitions. 

\begin{figure*}[t]
\centering
\includegraphics[width=1.0\columnwidth]{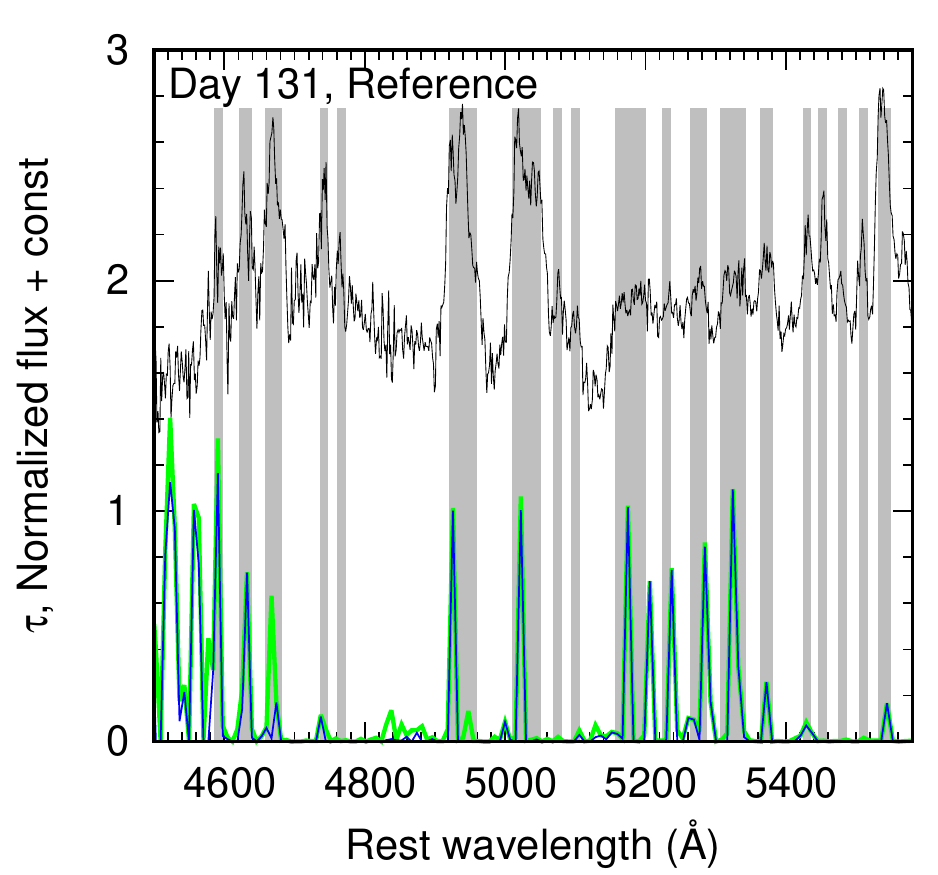}
\includegraphics[width=1.0\columnwidth]{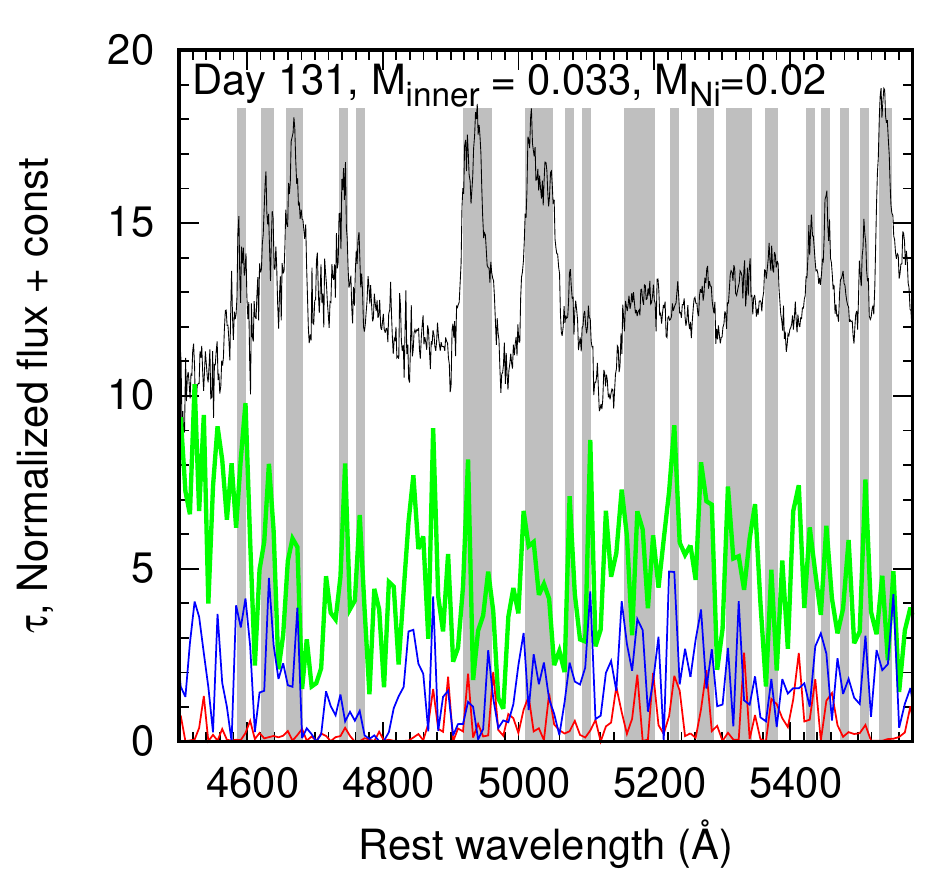}
\caption{An expanded view on the spectral region between 4,500 and 5,600\AA\ on day 131, as compared to the optical-depth distribution of the reference model (left) and that of the $^{56}$Ni-rich inner core having the total mass of $0.033 M_\odot$ and the $^{56}$Ni mass of $0.02 M_\odot$ (right). The optical depth is shown in a linear and absolute scale (Fe II shown by blue; Fe I by red; total including other species by green), while the observed spectrum is shown with different normalization in the two panels. Strong lines in the observed spectrum are marked by the gray-shaded areas. 
}
\label{fig:line_tau_expanded}
\end{figure*}

\subsection{Analyses of Allowed lines}

To further identify the reason(s) for the discrepancy between the reference model and the observed late-time spectra, we perform additional analyses in this section. This should help to understand the conditions in the innermost region which is indeed beyond the applicability of the reference model constructed through modeling the early-phase spectra.

We compute the Sobolev optical depths for various lines in the expansion opacity formalism at different wavelengths (with the wavelength range for the integration corresponding to $\sim 500$ km s$^{-1}$), under the LTE assumption for the ionization and excitation, with the (homogeneous) density and electron temperature given as the input parameters. The line list includes $\sim 5 \times 10^5$ transitions from \citet{kurucz1995}. The treatment of the expansion opacity can be found in, e.g., \citet{maeda2014}. Fig. \ref{fig:onezone_ref} shows the results for the model which mimics the reference ejecta model; the central density is set at the central density of the reference model following the homologous expansion, and the temperature is set to be $6,000$K (day 131) or $4,000$K (day 480) as guided by the results of the TARDIS calculations (Sections 4.1 and 4.3). Fig. \ref{fig:line_tau_expanded} shows an expanded view at $\sim 5,000$\AA\ (for demonstration) with the optical-depth distribution shown in a linear scale, for the easier comparison between the observed spectral features and the model optical-depth distribution. 

Fig. \ref{fig:onezone_ref} clearly explains the results of the TARDIS spectral synthesis calculations performed for the reference model (Section 4.1). On day 131, below $\sim 5,500$\AA, the optical depth is close to unity in some wavelength ranges, and the pattern indeed matches to the wavelengths in the observed spectrum where strong spectral-line features are found (see also the left panel of Fig. \ref{fig:line_tau_expanded}). We find that the features are largely contributed by Fe II (see also Fig. \ref{fig:fe_profile}). 
However, Fig. \ref{fig:line_tau_expanded} also shows that there are some strong lines in the observed spectrum which do not have corresponding lines in the model optical-depth distribution. Indeed, in the red portion of the spectrum ($\gsim 5,500$\AA), strong lines are rarely seen in this model. 
On day 480, the ejecta are largely transparent to any allowed transitions, and thus no or few strong features are expected. These properties explain the synthesized spectra shown in Fig. \ref{fig:model_ref}.

The qualitative match in the wavelengths of the strong features at $\lsim 5,500$\AA\ between the model and the observed spectrum strongly indicates that Fe II transitions provide a major contribution to the observed spectral-line features on day 131. This is consistent with the previous works suggesting large contributions by Fe II transitions in late-time spectra of SNe Iax \citep{jha2006,sahu2008,mccully2014b}. 
However, some lines are missing or too weak in the model below $\lsim 5,500$\AA\ (Fig. \ref{fig:line_tau_expanded}) and little line features are expected at $\gsim 5,500$\AA. In the reference model considered here, Fe as mostly produced by the decay chain $^{56}$Ni/Co/Fe occupies a large fraction of the ejecta ($\sim 40$\%). In the model, the ionization fraction of Fe$^{+}$ is $\sim 20$\% on day 131, and $\sim 100$\% on day 480. Therefore, there is no way to substantially increase the optical depths for Fe II transitions by changing the ionization status; even if Fe$^{+}$ dominates the ionization stage on day 131, this will never result in spectral features as strong as observed, given that the optical depths in most of the wavelength range at $\gsim 5,500$\AA\ are less than 0.01.

\begin{figure*}[t]
\centering
\includegraphics[width=1.05\columnwidth]{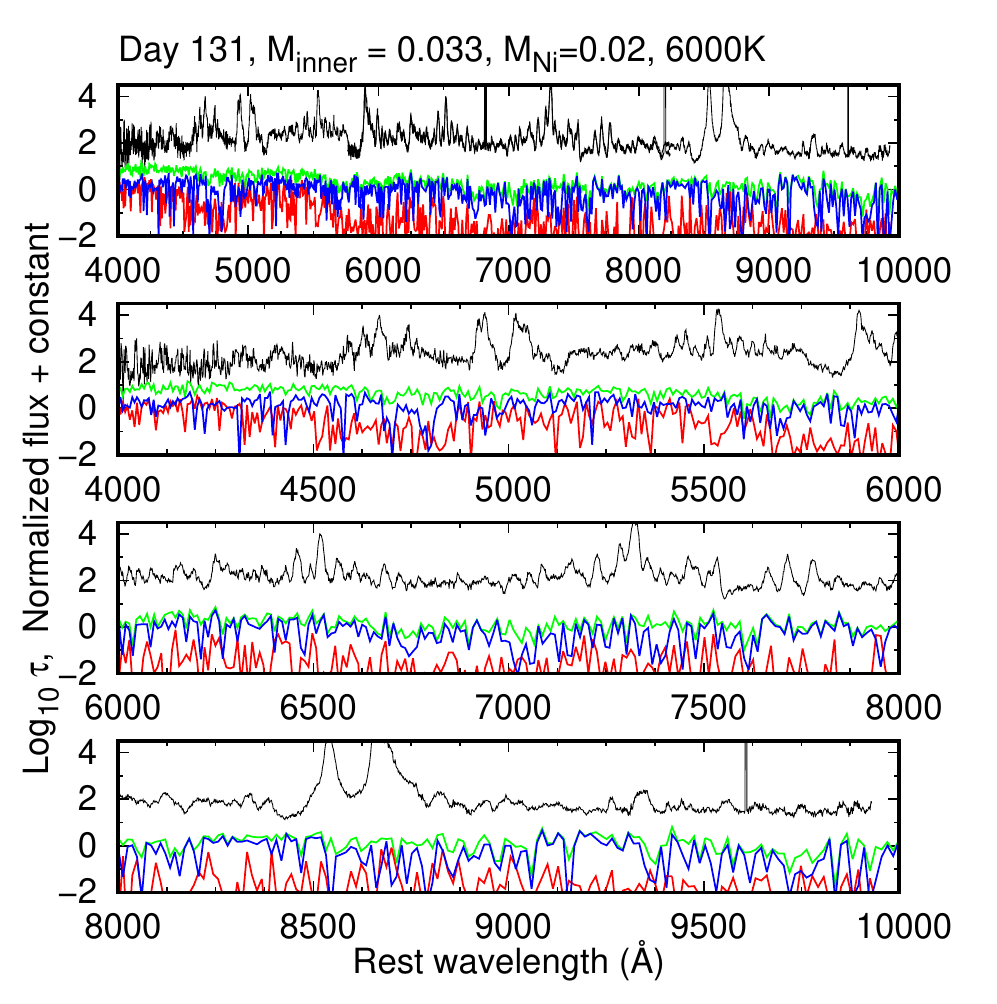}
\includegraphics[width=1.05\columnwidth]{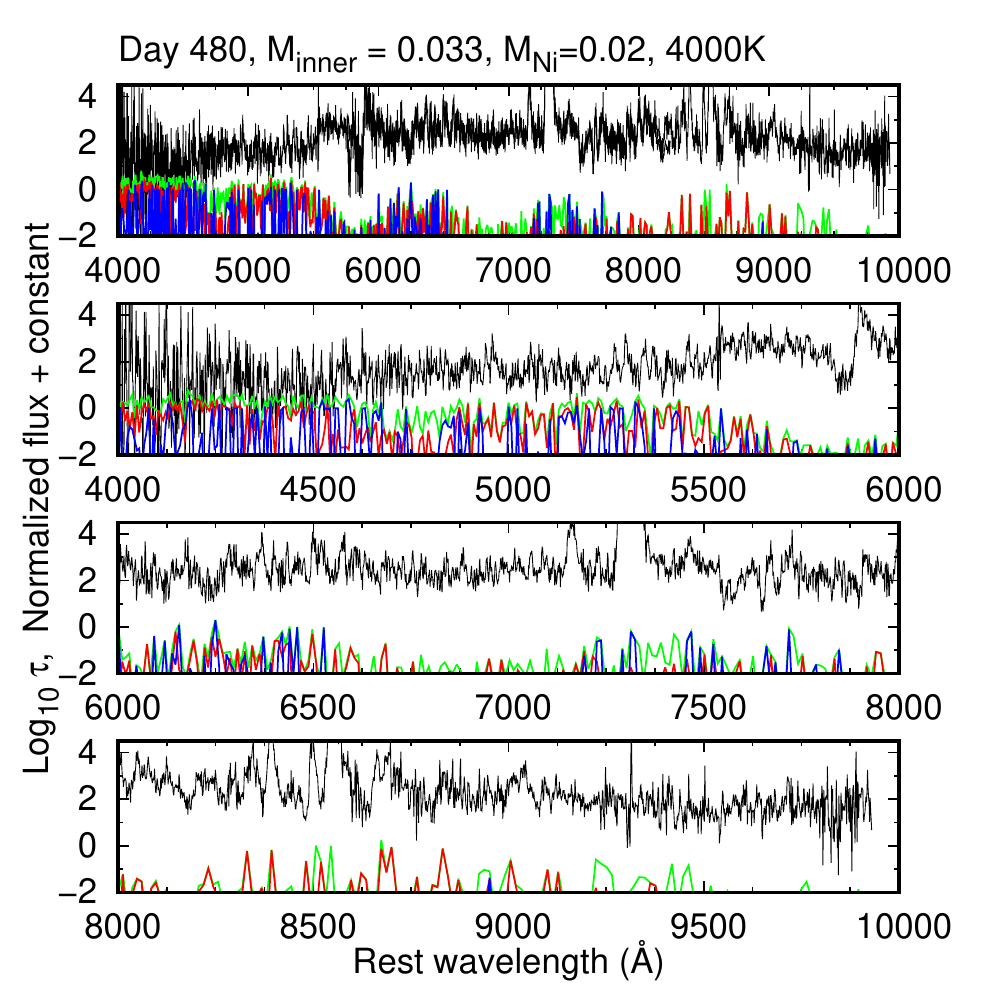}
\includegraphics[width=1.05\columnwidth]{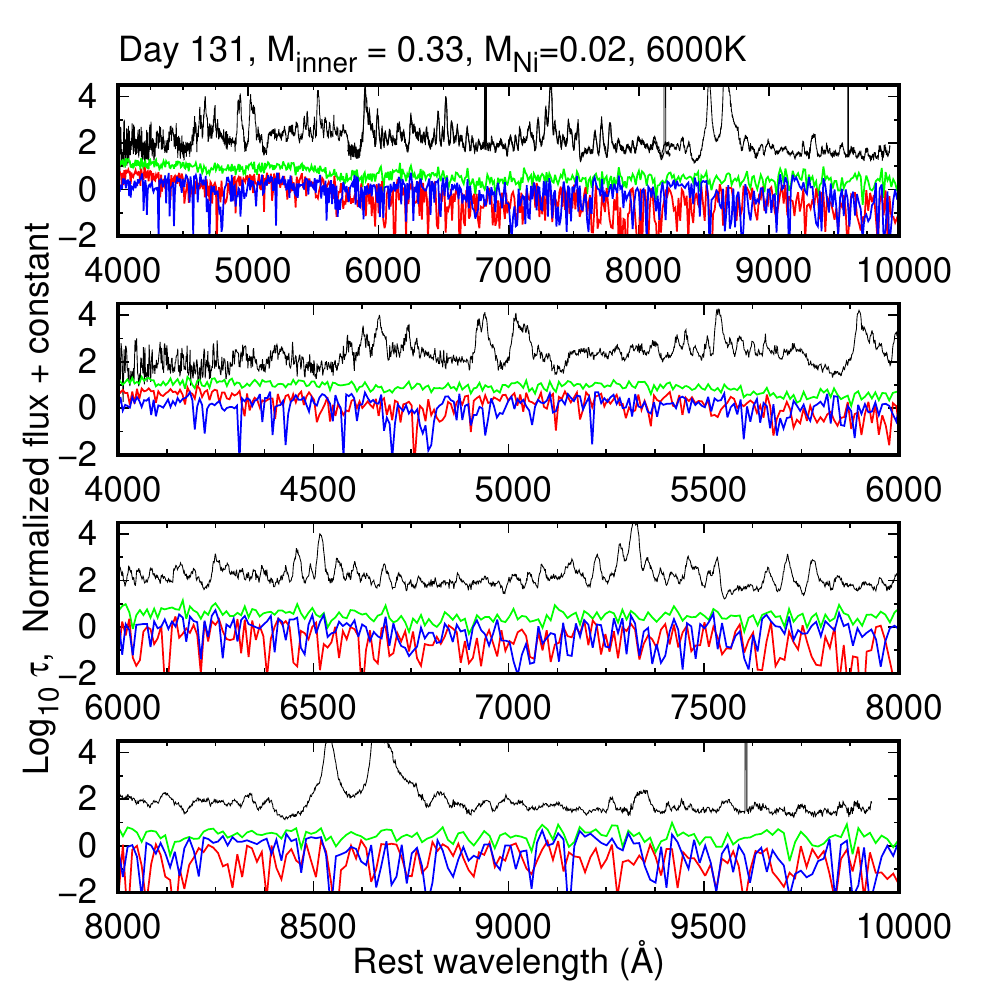}
\includegraphics[width=1.05\columnwidth]{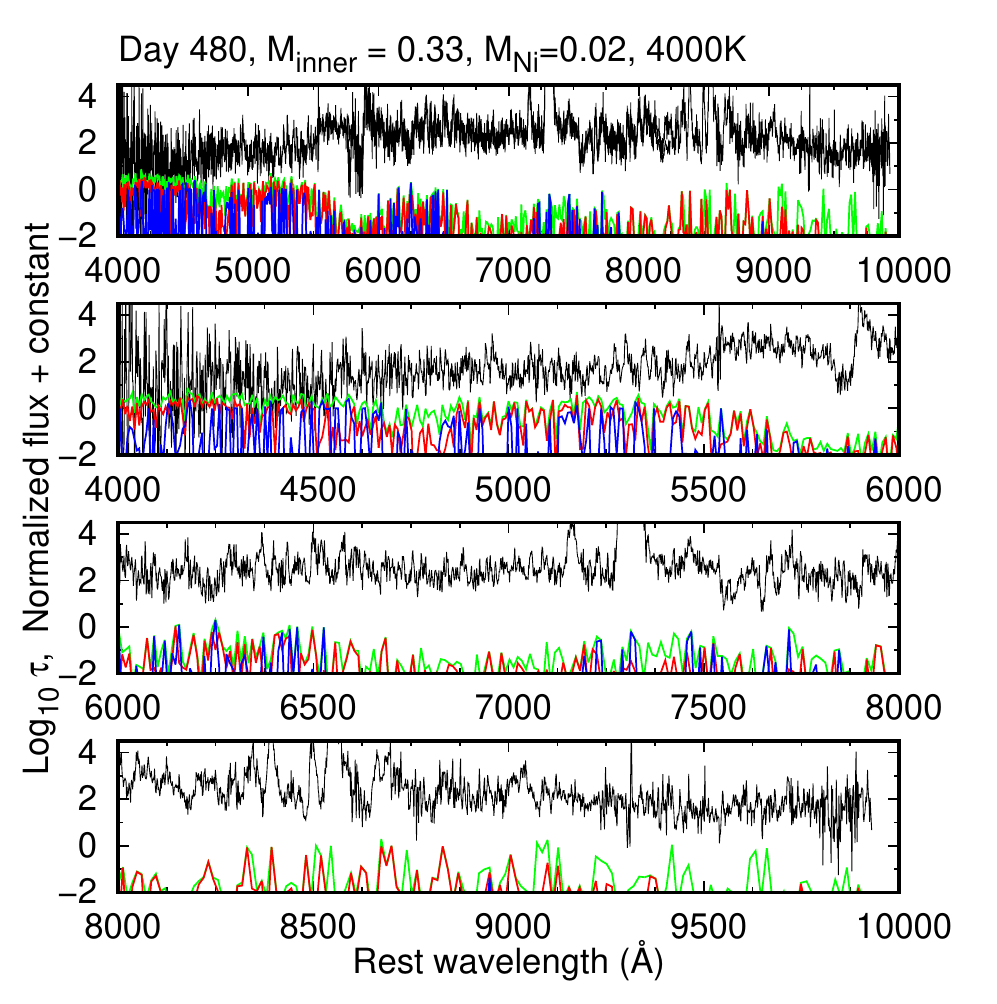}
\caption{The same as Fig. \ref{fig:onezone_ref}, but for the $^{56}$Ni-rich inner core having the total mass of $0.033 M_\odot$ and the $^{56}$Ni mass of $0.02 M_\odot$, without C nor O (upper panels). The same but for the C+O-rich inner core having the total mass of $0.33 M_\odot$ is shown in the lower panels, where the $^{56}$Ni mass is $0.02 M_\odot$ (therefore this model is rich in C and O). 
}
\label{fig:onezone_final}
\end{figure*}

Through this exercise, we robustly conclude that the inner density in the reference model is too low. This is not inconsistent with the conclusions by \citet{barna2021}, since the reference model is applicable only at $\gsim 3,500$ km s$^{-1}$. The main aim here is to further constrain the structure of the innermost region, to see if the inner structure is simply represented by smooth extrapolation from the outer layers or there is indeed another (distinct) component whose origin can be different from that of the outer ejecta. In other word, in the context of the failed delegation scenario that can explain main features of early-phase observational data by the unbound ejecta for SN 2019muj \citep{barna2021,kawabata2021} and other SNe Iax \citep{magee2016,barna2017,barna2018,srivastav2020, magee2022}, we want to test if there exists an additional component in the inner region. 

Indeed, some previous works point to the existence of a high-density inner region. First of all, the allowed-line dominated, photospheric spectra qualitatively indicate the high density \cite[e.g.,][]{jha2006}. Analyses of the line-flux ratio between [Ca II]$\lambda\lambda$7292, 7324 and the Ca II NIR triplet also indicate the high density \citep{sahu2008,stritzinger2015,mccully2014b}. 
The analysis of the late-time light curve of SNe Iax shows that the predicted luminosities from the unbound ejecta are not sufficient to explain the late-time luminosities of SNe Iax \citep{kawabata2018,kawabata2021}. Especially, \citet{kawabata2021} suggested that a combination of the (slightly-modified) unbound ejecta of the model N1def of \citet{fink2014} and an additional inner component with $M$($^{56}$Ni) $\sim 0.018 M_\odot$ can reproduce the late-time light curve behavior of SN 2019muj. 

Guided by the spectral analysis in the present work and the light curve analysis in the previous work, we consider an inner component having $\sim 0.02 M_\odot$ of $^{56}$Ni. Further insight can be obtained from the results of the line-profile study (Section 3.2); it also suggests an existence of a distinct inner component with the characteristic velocity of $\sim 760$ km s$^{-1}$. The most important element is likely Fe (and $^{56}$Ni as the parent nucleus), and therefore the detail of the composition content within the inner core is probably not important in studying the formation of the allowed transitions 
as long as the mass of Fe is fixed. In the present work, we take the characteristic composition pattern of the `bound' WD remnant in the failed deflagration scenario (see Section 5 for the possible interpretation). Specifically, we test two cases; (1)  the `$^{56}$Ni-rich composition' assumes that the inner component consists of Si, S, Ca, Ti, Fe, Co, Ni, $^{56}$Ni with relative fractions taken from the model N1def, and (2) the `C+O-rich composition' adopts the masses of Si -- Ni as same with case (1) but the total mass is increased by a factor of 10 as attributed to C and O (with equal mass fractions); namely, in case (2), the fraction of $10$\% is the Si -- Ni composition and the remaining 90\% is the C+O composition. The mass of $^{56}$Ni is fixed as $0.02 M_\odot$ in both cases. The total masses are $0.033 M_\odot$ and $0.33 M_\odot$, respectively, for cases (1) and (2). 

With the characteristic velocity of $\sim 760$ km s$^{-1}$, we can derive the average density of the inner component with the total mass of each model as mentioned above. This is $2.6 \times 10^{-14}$ g cm$^{-3}$ and $5.2 \times 10^{-16}$ g cm$^{-3}$, respectively, on days 131 and 480 for the $^{56}$Ni-rich ejecta ($0.033 M_\odot$). However, the average density is very likely an underestimate of the density at the line-forming region if it is substantially deep as compared to the characteristic velocity; for the exponential distribution with the e-folding radius $R_{\rm e}$, the central density is larger than the density of the homogeneous sphere with the outer radius set as $R_{\rm e}$, by a factor of 6. This effect is taken into account in the characteristic density adopted for the inner component in the following analyses; $1.6 \times 10^{-13}$ g cm$^{-3}$ and $3.1 \times 10^{-15}$ g cm$^{-3}$, respectively, on days 131 and 480, for the $^{56}$Ni-rich case ($0.033 M_\odot$). The density here is larger than the central density in the reference model by a factor of $\sim 400$. For the C+O-rich case ($0.33 M_\odot$), the density is increased further by an order of magnitude. 

With these conditions, we perform the same exercise of computing the optical depths as we have done for the reference model, using the Sobolev opacity and the expansion opacity formalism. The temperature is set to be the same with the case for the reference model; $6,000$ K on day 131 and $4,000$ K on day 480. Note that the calculations here are for a demonstration purpose; we will perform the spectral synthesis calculations in the next section. Fig. \ref{fig:onezone_final} shows the outcomes of these calculations for the $^{56}$Ni-rich inner component (upper panels) and the C+O-rich inner component (lower panels). 

The upper panels of Fig. \ref{fig:onezone_final} show that the $^{56}$Ni-rich model provides many allowed transitions that can contribute to the spectral features. On day 131, nearly the entire optical wavelengths have the optical depths to the allowed transitions exceeding or around unity. This explains the formation of the `photospheric' spectrum rather than the `nebular' spectrum even in such a late phase. Fe II transitions provide major contributions everywhere, and it is seen that the pattern of the optical depth distribution shows a good match to the observed spectrum, strengthening that the observed spectral features are mainly contributed by Fe II on day 131; this is further demonstrated in Fig. \ref{fig:line_tau_expanded} for the wavelength range between 4,500 and 5,600\AA. We note that there are some lines expected to be strong in the model optical-depth distribution, which are however not strong in the observed spectrum. We however emphasize that the detailed match, including the relative strengths of the spectral features, is beyond the scope of the exercise here, as it must take into account the radiation transfer effect. For example, there is a cluster of strong lines in the model between $\sim 4,800$ and $4,900$\AA, while they are not very strong in the observed spectrum. These lines can indeed been killed by blue-shifted absorption component of strong lines seen at $\sim 4,900-5,000$\AA. This is essentially the same effect we have discussed on the absorption by Fe II 7310, 7312, 7323, which is clearly seen in the profile of [Ca II]$\lambda\lambda$7292, 7324 (Section 3.2).The radiation transfer effects will be taken into account in the spectral synthesis simulations presented in the next section; note that the purpose of the analyses in this section is to robustly address the following two points adopting the simplified approach including the most relevant (and minimal) physical process (i.e., optical depth) but omitting more complicated processes (e.g., radiation transfer and line absorptions); (1) Fe II ad Fe I have strong transitions corresponding to the most distinct features in the spectra, and (2) these liens can be formed only in the high-density condition. 

The optical depth decreases on day 480, but it is still overall high; it is generally above unity at $\lsim 5,000$\AA, and it is between 0.01 and 1 in many wavelength ranges at $\gsim 5,000$\AA. Interestingly, under the condition considered here, Fe I transitions become to produce the dominant contribution while Fe II transitions are still visible \citep[see also][]{mccully2014b}. Again, we find that the observed spectral features are reasonably well traced by the distribution of the expansion opacity. The ionization state is dominated by Fe$^{+}$ in both epochs. The ionization fraction of the neutral Fe on day 480 in the model is $\sim 0.02$\% despite its major contribution to the opacity.

\begin{figure}[t]
\centering
\includegraphics[width=1.0\columnwidth]{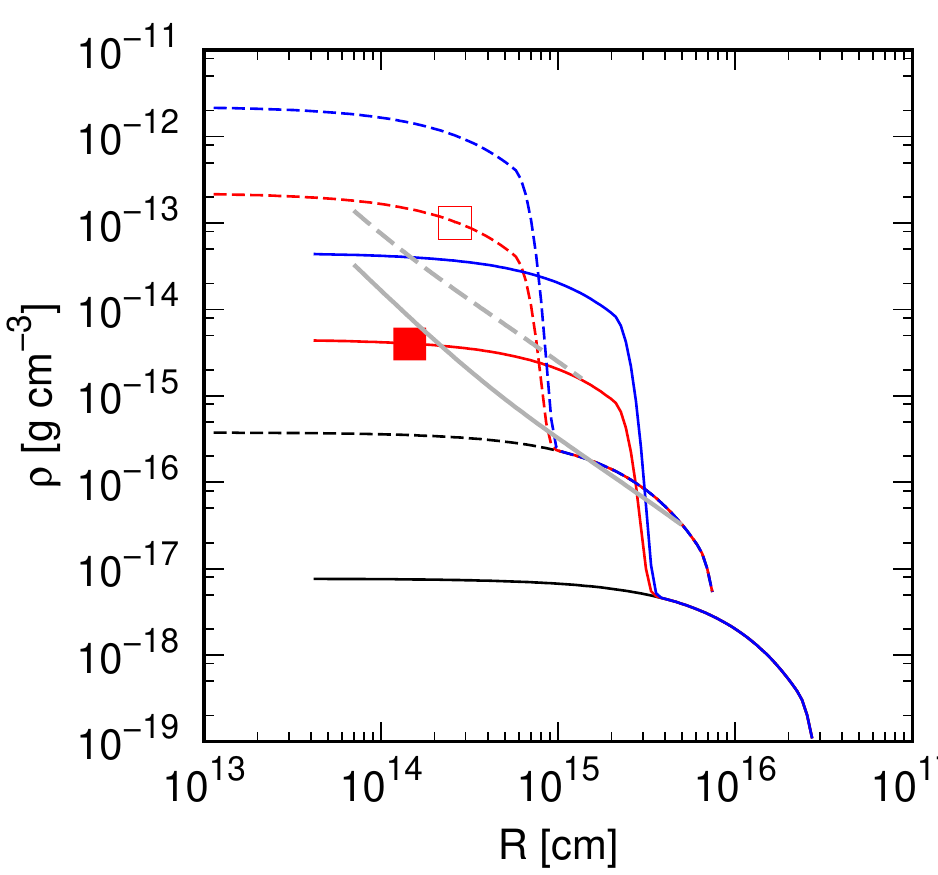}
\caption{The density distributions assumed in the TARDIS spectral synthesis calculations; the reference model (black), the $^{56}$Ni-rich inner core (red), and the C+O-rich inner core (blue).  The dashed lines are for day 131 and the solid lines are for day 480, assuming the homologous expansion. The squares mark the positions of the photosphere on day 131 (open) and day 480 (filled) overplotted on the $^{56}$Ni-rich model. Also plotted here for comparison are the outermost `homologous' component in the `wind' model of \citet{shen2017} scaled to days 131 and 480 (gray; see Section 5 for details).
}
\label{fig:density}
\end{figure}

Fig. \ref{fig:onezone_final} (lower panels) shows that the C+O-rich model leads to nearly identical optical-depth distribution with the $^{56}$Ni-rich model. We note that the mass of the $^{56}$Ni is the same between the two models, therefore the density of Fe is essentially the same. The main difference between the C+O-rich model and the $^{56}$Ni-rich model is seen in the stronger contribution of Fe I transitions on day 131 in the former. This stems from the additional electrons provided by C, which is mainly at C$^{+}$ for the model on day 131. On day 480, both C and O are mainly neutral, and therefore there is little difference between the two models. 

The exercises performed in this section show that a high density core, with $M$ ($^{56}$Ni) $\sim 0.02 M_\odot$, can explain the observed late-time spectra of SN 2019muj. It explains the emergence of the `photospheric' spectra rather than the `nebular' spectra. We have further clarified that the dominant transitions are Fe II on day 131 and Fe I on day 480. More detailed comparisons to the observed spectra require radiation transfer and spectral synthesis simulations (Section 4.3). Given the dominance of the Fe II and Fe I transitions, the content of other elements, e.g., whether it would indeed be dominated by the C+O composition, is difficult to determine solely from the arguments in this section; analyzing the forbidden lines is a key that provides additional constraints (Sections 4.4 and 4.5). 

\subsection{Spectral Synthesis: The Existence of the Inner Core}

\begin{figure*}[t]
\centering
\includegraphics[width=1.0\columnwidth]{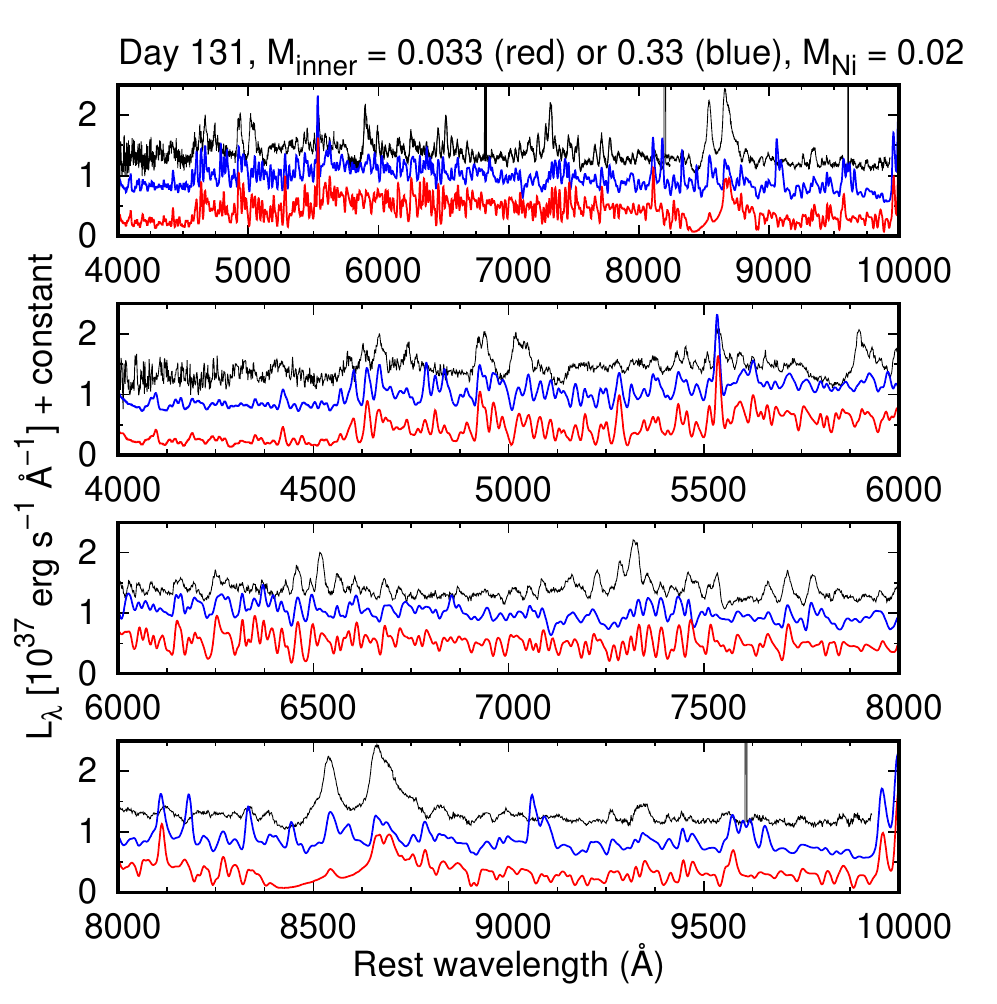}
\includegraphics[width=1.05\columnwidth]{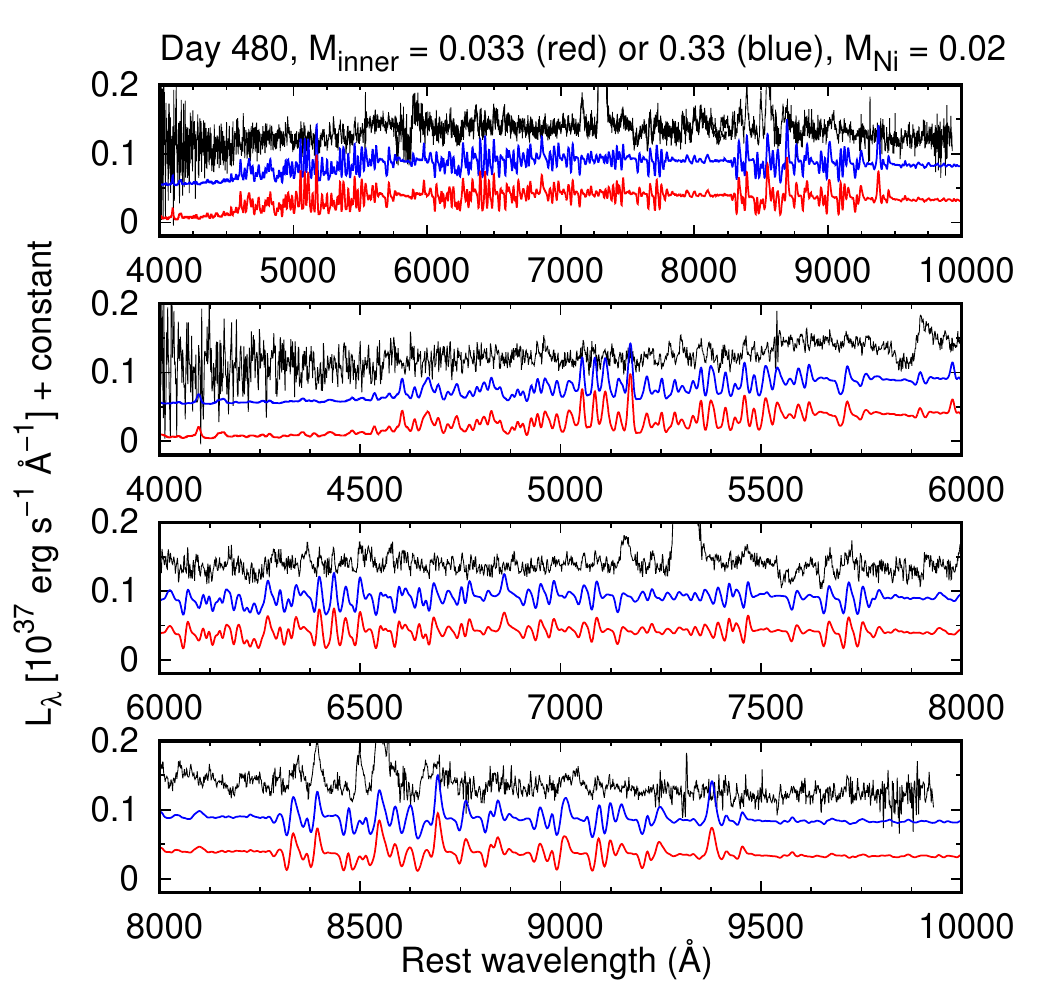}
\caption{The TARDIS spectral-synthesis models on days 131 (left) and 480 (right), for the density/abundance structure with the inner dense core; the $^{56}$Ni-rich compositions (red) and the C+O-rich compositions (blue). As the model input, the bolometric luminosity and the photospheric velocity are set as follows; $10^{7.05} L_\odot$ and $230$ km s$^{-1}$ on day 131, and $10^{6.05} L_\odot$ and $35$ km s$^{-1}$ on day 480. The photospheric temperature is 5800 K on day 131 and 4100 K on day 480 for the $^{56}$Ni-rich model; $\sim 7500$ K on day 131 and $\sim 4100$ K on day 480 for the C+O-rich model. 
}
\label{fig:model_combined}
\end{figure*}

The existence of the inner core, as argued in the previous sections by various analyses, is tested by the spectral synthesis calculations in this section. We add an additional inner component, with $M$ ($^{56}$Ni) $\sim 0.02 M_\odot$, to the reference model. The density distribution of the inner component is assumed to follow the same truncated exponential profile as the reference model but with different normalization and velocity scale; in the $^{56}$Ni-rich model, we adopt the central density of $\sim 5 \times 10^{-7}$ g cm$^{-3}$ $(t/{\rm day})^{-3}$ (which is larger than the corresponding density of the reference model by a factor of $\sim 580$), with the e-folding velocity of 300 km s$^{-1}$ and the truncated velocity of 500 km s$^{-1}$. The density of the inner component merges into the reference model at $\sim 750$ km s$^{-1}$. As in Section 4.2, we adopt the relative mass fractions of the bound WD remnant of model N1def for Si, S, Ca, Ti, Fe, Co, Ni and $^{56}$Ni for the inner component. As in the reference model, we have reduced the mass fractions of Ca and heavier elements above $1,000$ km s$^{-1}$. For $M$($^{56}$Ni) $\sim 0.02 M_\odot$, the total mass is $\sim 0.033 M_\odot$, i.e., dominated by $^{56}$Ni. In addition to the $^{56}$Ni-rich model, we compute the C+O-rich model, for which we increase the density further by a factor of 10 and this increase in the mass is attributed to C and O; the total mass in the model is then $0.33 M_\odot$ while $M$($^{56}$Ni) is fixed to be $0.02 M_\odot$, i.e., dominated by the C+O compositions. The density structures of these two models, as well as the reference model, are shown in Fig. \ref{fig:density}. 

Fig \ref{fig:model_combined} shows the synthetic spectra of these two models with the inner core, on day 131 and day 480. These two models show the spectral line features in the entire wavelength, not only on day 131 but on day 480. The patterns are formed largely through Fe II on day 131, as well as by Fe II and Fe I on day 480, which largely match to the observed spectral features. Therefore, we confirm the argument for the inner core through the spectral synthesis calculations. From the synthesized spectra, it is hard to conclude which of the $^{56}$Ni-rich composition or the C+O composition is a better solution. Namely, the mass of Fe (and $^{56}$Ni) is well constrained, but the amount of the other elements is not strongly constrained by this analysis. 

The inner core can also provide a possible explanation of the long-lasting photospheric phase (see Section 5 for further details). Fig. \ref{fig:photosphere} shows that the photosphere can be maintained in the late phase for the $^{56}$Ni-rich model, for a reasonable range of the opacity. If the same opacity ($0.025 - 0.5$ cm$^{2}$ g$^{-1}$, as appropriate for the Si-rich or Fe-rich compositions) is assumed, the C+O-rich inner core has too high density to be consistent with the photospheric velocity on days 131 and 480. A higher optical depth in the C+O-rich model is inferred from Fig. \ref{fig:onezone_final}, but this is only indicative; to address the issue further, we need to perform a complete radiation transfer calculation including the self-consistent formation of the photosphere.

\subsection{Analyses of [Ca II]$\lambda\lambda$7292, 7324}

In this section, we analyze [Ca II] $\lambda\lambda$7292, 7324, aiming at providing further constraints on the nature of the `inner component'. From the spectra, we measure the total flux of the [Ca II] (as a combination of the doublet) as $\sim 5 \times 10^{38}$ erg s$^{-1}$ on day 131 and $\sim 1 \times 10^{38}$ erg s$^{-1}$ on day 480. Given that (1) numerous allowed transitions are seen in the spectra, and that (2) the region emitting the forbidden lines are overlapping with the region processing the allowed transitions, it is likely that the temperature is determined by the heating and cooling through these allowed transitions, rather than the non-thermal heating and cooing via forbidden lines as usually associated with the forbidden-line formation in SNe. From the TARDIS simulations, we see that the electron temperature within $\sim 500$ km s$^{-1}$ is $\sim 6,000$ K on day 131 and $\sim 4,000$ K on day 480, and we use these values as references in the following discussion. 

If the density is sufficiently high so that the level populations of  4s$^2$S and 3d$^2$D, i.e., the lower and the upper levels for the [Ca II], can be described by the LTE, the line luminosity under the one-zone approximation is described as follows \citep[e.g.,][]{li1993}: 
\begin{eqnarray} 
& L ({\rm [Ca II]}) & \sim 5.4 \times 10^{40} \ {\rm erg s}^{-1}  \nonumber\\ & & P_{\rm esc} exp\left(\frac{-19,700 {\rm K}}{T_{\rm e}}\right) \left(\frac{M ({\rm Ca}^{+})}{10^{-4} M_\odot}\right)\ ,
\end{eqnarray}
where $T_{\rm e}$ is the electron temperature, $M({\rm Ca}^{+})$ is the mass of Ca$^{+}$ in the emitting region, and $P_{\rm esc}$ is the escape probability of the line(s). Inserting the temperatures ($6000$ K on day 131 and $4000$ K on day 480), the exponential (Boltzmann) factor decreases from day 131 to day 480 by a factor of $\sim 5$. This is the expectation in the optically-thin line-luminosity evolution in case 
the emitting regions in the two epochs are identical. This evolution is quite consistent with the observed luminosity decrease between the two epochs, indicating that this simple picture probably grabs the basic mechanism of the [Ca II] formation in SN 2019muj. 

To evaluate the uncertainty, we may consider a range of the electron temperature. We conservatively assume that the electron temperature is in the range of $5,000 - 7,000$ K on day 131 and $3,500 - 4,500$ K on day 480. This roughly covers a range of the differences we see in the TARDIS simulations for the reference model and the two models with the ($^{56}$Ni-rich or C+O-rich) inner core (Section 4.3). 

From equation (1), we can estimate the mass of Ca$^{+}$ ions. On day 131, we derive $M$ (Ca$^{+}$) $\sim 4.7 \times 10^{-5} M_\odot$ (for 5000 K), $\sim 2.5 \times 10^{-5} M_\odot$ (6000 K), or $\sim 1.5 \times 10^{-5} M_\odot$ (7000 K); on day 480, $M$ (Ca$^{+}$) $\sim 5.0 \times 10^{-5} M_\odot$ (for 3500 K), $\sim 2.5 \times 10^{-5} M_\odot$ (4000 K), or $\sim 1.5 \times 10^{-5} M_\odot$ (4500 K). 

We thus constrain the Ca mass in the inner ejecta below $\sim 760$ km s$^{-1}$ to be $\gsim (1-5) \times 10^{-5} M_\odot$. The conversion of the mass of Ca$^{+}$ to the total mass of Ca is simpler on day 480 than on day 131; for the temperature found for day 480, there is no doubt that Ca$^{+}$ dominates the Ca ionization stage. We therefore derive $M$ (Ca) $\sim (1-5) \times 10^{-5} M_\odot$. On the other hand, the ionization matters on day 131. For the $^{56}$Ni-rich one-zone model, the dominant ionization stage for Ca at $6000$ K is indeed the doubly-ionized ion with only $\sim 2$\% of Ca in the singly-ionized state. This will require a large amount of Ca ($\gsim 10^{-3} M_\odot$) to explain the derived mass of Ca$^{+}$. On the other hand, if the temperature would be 5000 K, Ca$^{+}$ dominates the ionization stage; the total Ca mass in this case is then $\sim 6 \times 10^{-5} M_\odot$. In view of the rough consistency to the value derived on day 131, the latter is more likely the case, and we suggest it is most likely that $M$ (Ca) $\sim 5 \times 10^{-5} M_\odot$. 

The LTE requires that the electron density ($n_{\rm e}$) is above the critical density $n_{\rm cr}$ \citep[][]{li1993}: 
\begin{equation}
n_{\rm e} \gsim n_{\rm cr} \sim 7.3 \times 10^{5} \left(\frac{T_{\rm e}}{1,000 \ {\rm K}}\right)^{1/2} \ {\rm cm}^{-3} \ .
\end{equation}
For the above mentioned model(s), we derive $n_{\rm e} \sim (1-3) \times 10^{9}$ cm$^{-3}$ on day 131, and $n_{\rm e} \sim (2-6) \times 10^{7}$ cm$^{-3}$ on day 480. The LTE assumption is therefore well justified. 

Another factor that enters into the Ca mass estimate is the Sobolev optical depth ($\tau$) for the [Ca II]. The escape probability (for the one-zone model) is described as follows: 
\begin{equation}
P_{\rm esc} =\frac{1 - e^{-\tau}}{\tau} \ , 
\end{equation}
with the Sobolev optical depth estimated as follows \citep{li1993}: 
\begin{eqnarray}
& \tau & \sim 0.062 f_{\rm Ca}^{-1} \nonumber\\
& & \left(\frac{M ({\rm Ca}^{+})}{10^{-4} M_\odot}\right) \left(\frac{V}{2500 \ {\rm km} \ {\rm s}^{-1}}\right)^{-3} \left(\frac{t}{100 \ {\rm day}}\right)^{-2} \ , 
\end{eqnarray}
where $f_{\rm Ca}$ is the fraction of the volume occupied by the material emitting the [Ca II]. In the present situation, it is then computed as $\tau \sim 0.3 f_{\rm Ca}^{-1}$ on day 131 and $\sim 0.02 f_{\rm Ca}^{-1}$ on day 480, for $M$ (Ca$^{+}$) $\sim 2 \times 10^{-5} M_\odot$ and $V \sim 760$ km s$^{-1}$. In the simplest case of the fully (microscopic) mixing, $f_{\rm Ca}$ is unity, and thus $\tau < 1$ on both epochs. As another possibility, we may take $f_{\rm Ca} \sim M_{\rm Si} / M_{\rm tot}$, where $M_{\rm Si}$ is the mass of the Si-rich material (as a main formation site for Ca) and $M_{\rm tot}$ is the total mass. This formalism corresponds to the opposite extreme where the region emitting the [Ca II] is macroscopically separated from the other region \citep[e.g.,][]{maeda2007,dessart2020}; for the WD remnant of the N1def model but omitting the C+O-rich composition (i.e., the $^{56}$Ni-rich inner component), this is described as $f_{Ca} \sim 0.25$. Within the uncertainty, we therefore see that the optical depth is at most around unity. This roughly justifies our assumption, but may increase the total Ca mass by a factor of a few. 

If the total mass of the inner ejecta below $\sim 760$ km s$^{-1}$ is $0.033 M_\odot$, the (average) mass fraction of Ca is $\sim 1.5 \times 10^{-3}$. This may be further increased by a factor of a few by the optical depth effect, if the [Ca II]-emitting region (i.e., the Si-rich region) is macroscopically separated from the other regions (see above). We note that the Ca mass fractions found both in the bound remnant and unbound ejecta of the model N1def of \citet{fink2014} are $\sim 0.01$, which is roughly consistent with the value found in the present analysis (see Section 5 for details). This finding may be regarded as an additional support for the failed thermonuclear explosion scenario for SN 2019muj (and SNe Iax in general). 

\subsection{Constraints on the oxygen content}

An important remaining question is weather the inner component is indeed dominated by the C+O composition, which has not been strongly constrained by the analyses in the previous sections. In this section, we address this issue by examining the condition for the formation of [ O I]$\lambda\lambda$6300, 6363 to satisfy the upper limit on its strength. 

Given the possible detection of the [O I], we start with the day 480 spectrum. Similarly to the analysis of the [Ca II], the [O I] luminosity for the 6300 component is described as follows under the LTE condition \citep[e.g.,][]{uomoto1986}.; 
\begin{eqnarray}
& L ({\rm [O I]}) & \sim 1.5 \times 10^{42} \ {\rm erg s}^{-1}  \nonumber\\ & & P_{\rm esc} exp\left(\frac{-22,860 {\rm K}}{T_{\rm e}}\right) \left(\frac{M ({\rm O}^{+0})}{M_\odot}\right)\ .
\end{eqnarray}
The flux normalization is smaller for the 6363 component by a factor of 3. 

Putting the flux of the possible [O I]$\lambda$6300 on day 480 (which can also be regarded as an upper limit) assuming $T_{\rm e} \sim 4,000$ K, we obtain $M_{{\rm O}^{+0}}\lsim 1.5 \times 10^{-3} M_\odot$. The Sobolev optical depth is scaled as follow: 
\begin{eqnarray}
& \tau & \sim 17 f_{\rm O}^{-1} \nonumber\\
& & \left(\frac{M ({\rm O}^{+0})}{M_\odot}\right) \left(\frac{V}{1000 \ {\rm km} \ {\rm s}^{-1}}\right)^{-3} \left(\frac{t}{100 \ {\rm day}}\right)^{-2} \ , 
\end{eqnarray}
where $f_{\rm O}$ is the fraction of the volume occupied by the [O I]-emitting region. We thus estimate $\tau \sim 2.6 \times 10^{-3} f_{\rm O}^{-1}$. Therefore, the optically-thin limit is applicable, unless a large amount of the [O I] emitting region is confined in a very small volume, a situation which would never be realized under the observational constraints for SN 2019muj. We may assume that $f_{\rm O}$ is represented by the relative fraction of the mass of the O-rich region to the total mass (i.e., the `macroscopic' mixing; Section 4.4). Then, as one extreme, we can assume that the mass of the [O I] emitting region is negligible as compared to the total mass. In this case, adopting the total mass of $\sim 0.033 M_\odot$, then $f_{\rm O} \sim 0.08$, leading to $\tau \sim 0.03$. As the other extreme, we may assume that the O-rich region dominates the mass of the inner component. In this case, $f_{\rm O} \sim 1$ by definition, and therefore it is inevitable to have $\tau \ll 1$. 

Indeed, the constraint can be generalized based on equations 5 and 6. In the optically-thin limit, $L ({\rm [O I]}) \propto M ({\rm O}^{+0})$ for given temperature. As the optical depth increases, the luminosity approaches asymptotically to $\sim 3 \times 10^{39}$ erg s$^{-1}$ for $T_{\rm e} \sim 4,000$ K as is independent from the mass. The change in the behavior takes place at $M ({\rm O}^{+0}) \sim 0.6 M_\odot$, corresponding to  $\tau \sim 1$, with the expected luminosity of [O I]$\lambda$6300 far exceeding the observational constraint ($\lsim 7.6 \times 10^{36}$ erg s$^{-1}$). As such, the upper limit we derive under the optically-thin limit, i.e., $M ({\rm O}^{+0}) \lsim 1.5 \times 10^{-3} M_\odot$, is robust, as long as the temperature is $\sim 4,000$ K. 

The electron density computed in the one-zone models for day 480 is $\sim 10^{7}$ cm$^{-3}$. Given the critical density of $\sim 3.6 \times 10^{6}$ cm$^{-3}$ for [O I] 6300, the LTE is roughly applicable. Also, at this condition, the oxygen ionization is dominated by the neutral state. We therefore place an upper limit of $\sim 1.5 \times 10^{-3} M_\odot$ for the oxygen content. 

A similar analysis is possible for the day 131 spectrum. Here, we derive the upper limit of oxygen as $\lsim  9 \times 10^{-4} M_\odot$, which is similar to the value obtained based on the day 480 spectrum. 

The above estimate is based on the assumption that the [O I]-emitting region, if they would exist, has the thermal condition similar to that in the region emitting the allowed transitions. It is possible that the O-rich region and other regions (e.g., $^{56}$Ni-rich region) are macroscopically separated. If the electron temperature would be as low as $\sim 2,000$K on day 480, $\sim 0.5 M_\odot$ of oxygen can be accommodated within the upper limit for the [O I]. To test this possibility, we have run an additional TARDIS model for day 480, using the same density structure of the C+O-rich model but adopting a pure C+O composition contaminated just by the solar abundance for the heavier elements. The resulting temperature structure in the model is not very different from the C+O-rich model (and the $^{56}$Ni-rich model), characterized by $\sim 4,000$ K. In the model, we see that the spectrum above $\sim 5,000$\AA\ is featureless except for the Ca II NIR and a few other lines, as is similar to the result of the reference model. Below $\sim 5,000$\AA\ and in the UV, the spectral lines are still seen, which controls the heating rate in the model. We also note that the observed limit for the [O I] luminosity is negligible as compared to the total luminosity, which means that the [O I] is not a dominant coolant. Therefore, it is likely that even if there is a C+O-rich region macroscopically separated from the $^{56}$Ni-rich region, the temperature is likely controlled by the heating-cooling balance through the allowed transitions. The results of the TARDIS calculations then indicate that the temperature in such a region will not be very different from the $^{56}$Ni-rich region. We therefore conclude that the upper limit we have derived for the oxygen content would not be altered even by considering the macroscopically-separated distribution between chemically different regions. 

\section{Delayed Radioactive Heating as a possible Origin of the Inner Core}

\begin{figure*}[t]
\centering
\includegraphics[width=1.5\columnwidth]{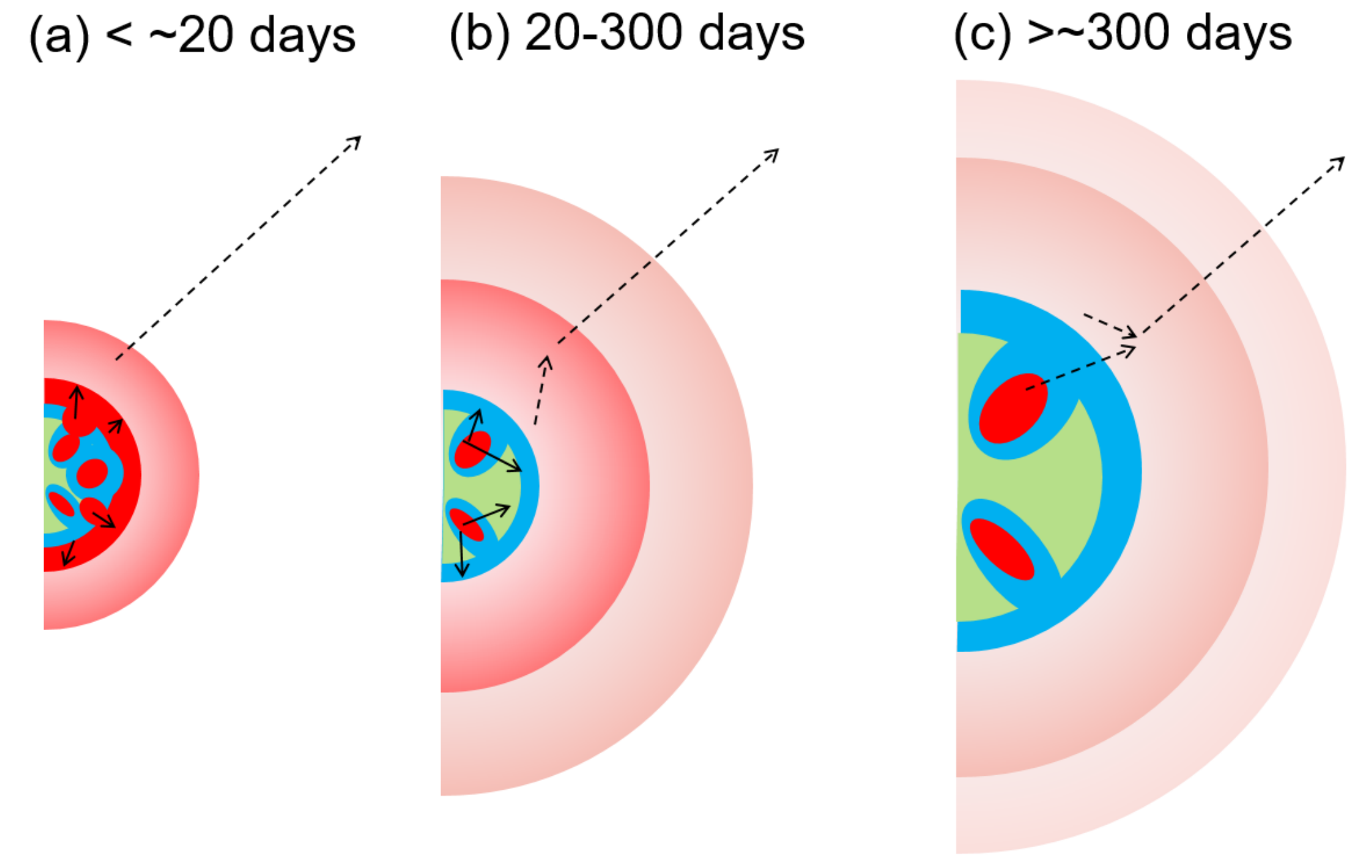}
\caption{A schematic picture of the proposed scenario. The $^{56}$Ni-rich, Si-rich (including Ca), and C+O-rich materials are (schematically) shown by red, blue, and green, respectively. (a) As the outcome of the initial thermonuclear explosion, $^{56}$Ni-rich materials are ejected as the (first) unbound ejecta. The radioactive decay of $^{56}$Ni/Co/Fe chain within the bound WD remnant keeps energizing the outer region of the bound WD (or the envelope; indicated by solid arrows). (b) The $^{56}$Ni/Co heating creates the `second' unbound ejecta (i.e., the inner component) in about a month. By this time, the outer ejecta become transparent, and the second ejecta controls the optical output (indicated by dashed arrows). The $^{56}$Ni/Co left deep within the WD keeps energizing the bound WD further. (c) The bound WD has further puffed up. When the photons start diffusing out even from the bound WD in about one year after the initial explosion, the expansion is virtually stopped. The photons originally emitted from the bound WD start contributing to the optical output. 
}
\label{fig:schematic}
\end{figure*}

Based on the analyses of the late-phase spectra of SN 2019muj, we conclude that there exists a dense inner component which is distinct from the outer ejecta. The nature of this inner component is strongly constrained; the mass of the innermost component is estimated to be $\sim 0.03 M_\odot$ and it is dominated by Fe (which can be initially $^{56}$Ni), expanding with the velocity of $\sim 760$ km s$^{-1}$. In this section, we discuss a possible origin of this inner component, especially in a context of the failed deflagration model of a (nearly $M_{\rm Ch}$) C+O WD. 

The characteristic properties of the inner component suggest a possible connection to the bound WD remnant. The bound WD remnant, especially in case that the mass of the unbound ejecta is small \citep[to be consistent with the early-phase properties of SN 2019muj;][]{barna2021,kawabata2021}, contains a large amount of $^{56}$Ni as is comparable to what is derived for the inner component. For example, in the model N1def of \citet{fink2014}, the mass of $^{56}$Ni left within the bound WD remnant is $\sim 0.033 M_\odot$ out of $\sim 1.33 M_\odot$ of the total WD remnant mass. The mass of the Ca in the bound WD remnant in the same model is $\sim 3 \times 10^{-4} M_\odot$, i.e., $\sim 1$\% of the mass of $^{56}$Ni. The mass of Ca we have derived for the inner component is $\sim 5 \times 10^{-5} M_\odot$ (with a possible underestimate by a factor of a few), which can be accommodated by the content of Ca in the WD remnant. We have a strong upper limit for the mass of O ($\lsim 1.5 \times 10^{-3} M_\odot$) which is far below the content within the WD remnant ($\sim 0.6 M_\odot$). 

The WD remnant predicted in the failed deflagration scenario can be a promising origin for the inner component, if there is a mechanism that can eject most of $^{56}$Ni, a large fraction of Ca (but with a smaller efficiency than the $^{56}$Ni ejection), and little C+O-rich materials. Below, we suggest that delayed radioactive input by the $^{56}$Ni/Co/Fe decay chain within the WD remnant could unbind a fraction of the bound WD remnant and create the `second' unbound component that follows the original unbound ejecta. This is schematically illustrated in Fig. \ref{fig:schematic}. This scenario is indeed similar (or even identical in many respects) to the suggestion by \citet{shen2017}, who examined the details of the evolution of the WD remnant by considering that a super-Eddington wind is launched from the WD remnant. In this section, we study the expected outcomes of the scenario in a phenomenological way, based on the energetic and radius/time scales associated with such a system. Such a phenomenological approach here is complementary to the discussion provided by \citet{shen2017}. Given that the mass ejection process can indeed be complicated and details can be dependent on the initial conditions (as an outcome of the thermonuclear explosion), we believe the discussion here can provide a useful basis to further investigate the details of the scenario. 

The WD remnant has probably puffed up following the thermonuclear runaway at least in its envelope structure, so the initial configuration expected as the aftermath of the failed deflagration scenario is a combination of a small amount of the unbound ejecta (for the least-energetic cases, i.e., for SN 2019muj) and the extended bound remnant. If the temperature within the remnant is not too high to fully ionize the radioactive isotopes \citep[but see][]{shen2017}, the radioactive decay starts providing additional energy input. The $^{56}$Ni-rich regions will then become hot bubbles, and will be accelerated toward the remnant WD surface. Once the deposited energy exceeds the binding energy of such materials, they will escape from the WD remnant, creating the `second' unbound ejecta. The energy input here takes place preferentially in the region dominated by $^{56}$Ni and therefore a large fraction of the $^{56}$Ni region in the bound WD remnant could be ejected during this process. The $^{56}$Ni bubble is expected to be initially located preferentially in the outer region of the bound WD (as this region must have been initially energized by the explosive nucleosynthesis, and also further driven outward by buoyancy); therefore, the ejection of the Si-rich region (including Ca) is probably less efficient, and the ejection of the C+O-rich materials will be very limited. This explains the compositions we have estimated for the inner component. 

If the remnant WD envelope has initially extended to $\sim 10^{10}$ cm as an outcome of the thermonuclear runaway \citep[if the hot envelope behaves in a similar manner to the binary WD merger;][]{dan2011,shen2012}, the binding energy of $\sim 0.033 M_\odot$ (to be ejected) on top of the WD core with $\sim 1.33 M_\odot$ (to remain bound) is $\sim 10^{48}$ erg. The comparable energy can be supplied by the decay chain $^{56}$Ni/Co/Fe in about two weeks with $M$($^{56}$Ni) $\sim 0.02 M_\odot$, and then the hot bubbles will start escaping the WD remnant. When an additional energy of $\sim 10^{47}$ erg is further deposited, the velocity of the second ejecta will reach to $\sim 760$ km s$^{-1}$. 

Given that the radioactive power (therefore the mass-loss driving force) decreases with time, the (second) unbound component once ejected well outside the left-over WD is expected to behave more like homologously expanding ejecta rather than the (steady-state) wind. We have confirmed this picture in Fig. 3 of \citet{shen2017}, where the velocity field at $\gsim 5 \times 10^{13}$ cm follows the homologous expansion law reasonably well. In the discussion below, we thus treat the `second' ejecta as homologously expanding materials. 

Once the $^{56}$Ni materials are ejected from the system, further energy input will be limited by the competition between the photon diffusion time scale and the expansion time scale. The diffusion time scale, adopting the opacity of $0.1$ cm$^{2}$ g$^{-1}$), is estimated as follows: 
\begin{equation}
\tau_{\rm diff} \sim 20 \ {\rm days} 
\left(\frac{M_{\rm ej}}{0.033 M_\odot}\right) 
\left(\frac{R_{\rm ej}}{10^{13} \ {\rm cm}}\right) \ ,
\end{equation}
where $M_{\rm ej}$ and $R_{\rm ej}$ are the mass and radius of the second ejecta (i.e., the inner component). Assuming the velocity of $760$ km s$^{-1}$, the photon diffusion becomes substantial in $\sim$ a week, after which further acceleration can be neglected and the photons from this inner component start contributing to the optical output. 

\begin{figure}[t]
\centering
\includegraphics[width=\columnwidth]{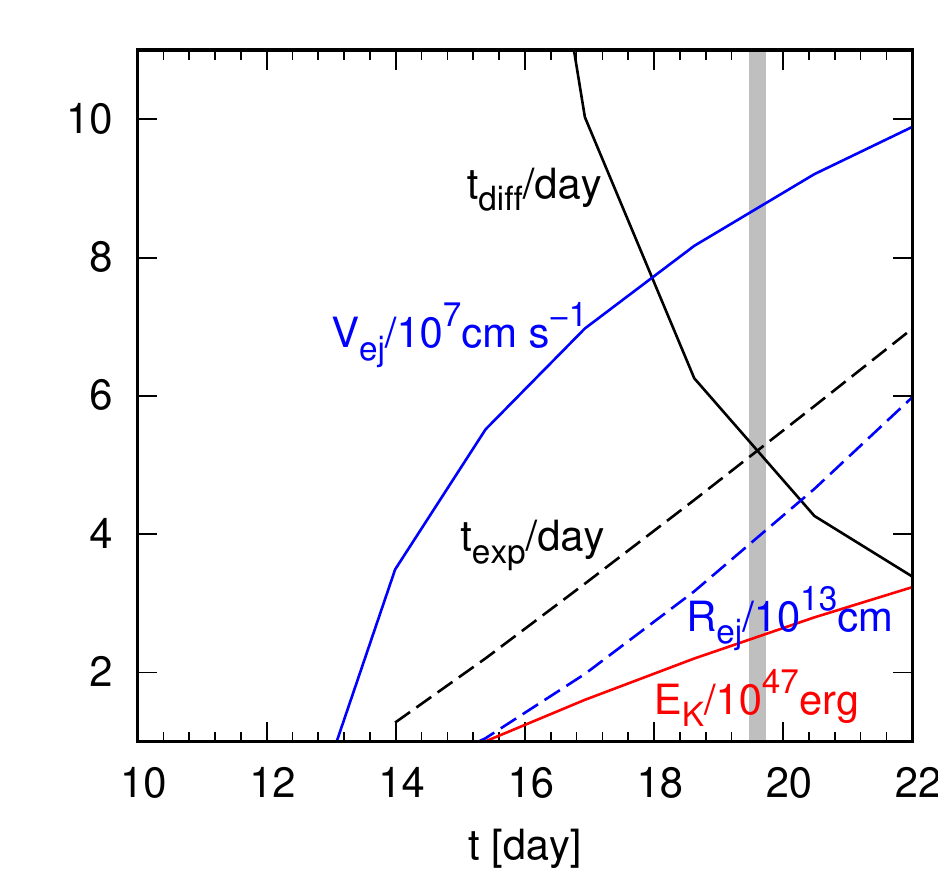}
\caption{A toy model for the evolution of the $^{56}$Ni shell (i.e., the inner component) which is initially within the WD remnant (or its envelope). Shown here are the kinetic energy (red), characteristic velocity (blue-solid), radius (blue-dashed), diffusion time scale (black-solid), and the expansion time scale (black-dashed). The vertical gray line shows the timing when the photon diffusion becomes important. 
}
\label{fig:wd_ej}
\end{figure}

Fig \ref{fig:wd_ej} shows the expected evolution of the $^{56}$Ni-rich shell (with the total mass of $0.033 M_\odot$ and the $^{56}$Ni mass of $0.02 M_\odot$). In this simple model, we assume that the $^{56}$Ni-rich shell is placed initially on top of a (puffed) WD with the mass and radius of $1.33 M_\odot$ and $10^{10}$cm, respectively. The radioactive energy generated within the $^{56}$Ni bubble is assumed to be all trapped. Once the deposited energy exceeds the binding energy of this shell ($\sim 1.2 \times 10^{48}$ erg), the shell is assumed to expand outward with the velocity corresponding to the kinetic energy gained by that time (i.e., the total deposited energy minus the binding energy). This ejection is initiated $\sim$ two weeks after the initial thermonuclear runaway. Initially the deposited energy is all trapped within the shell, and therefore the kinetic energy and velocity keep increasing (i.e., the shell is accelerated). As the expansion time scale ($t_{\rm exp} = R_{\rm ej}/V_{\rm ej}$) increases but the diffusion time decreases, at some point the photons start diffusing out of the shell. This situation takes place at $\sim 20$ days (or about a week after the shell ejection) when the shell reaches to $\sim 4 \times 10^{13}$ cm, after which the deposited energy will be efficiently released by photons and therefore the shell kinematic is frozen. We note that this simple picture seems to explain the key behaviors seen in the detailed simulations for the wind-driven ejection performed by \citet{shen2017}, where the photosphere is formed within the wind at $\sim 4 \times 10^{13}$ cm above which the velocity field roughly follows the homologous expansion law. 

In this simple model, we see that the final shell velocity is $\sim 900$ km s$^{-1}$, which is roughly consistent with $\sim 760$ km s$^{-1}$ we have derived for the inner component. This component is expected to start contributing to the optical output at $\sim 20$ days after the explosion, or $\sim 10$ days after the maximum light for SN 2019muj. This is indeed the timing when the light curve expected from the unbound ejecta starts under-reproducing the observed light curve of SN 2019muj and some other SNe Iax \citep{kawabata2021} (see below for further discussion on the light curve properties).

A fraction of $^{56}$Ni may be further left within the remnant WD. The additional power provided by the left-over $^{56}$Ni would not be sufficient to further create additional dynamical ejection. For example, if we assume $\sim 0.01 M_\odot$ of $^{56}$Ni is still left within the WD, the asymptotic energy generation is $\sim 2 \times 10^{48}$ erg, which is far below the binding energy of the WD remnant. We may then consider that the excessive energy is used for (slow) expansion of the WD remnant, for which the characteristic time scale may be roughly estimated by the dynamical time scale. By fixing the WD mass as $1.33 M_\odot$ and relating the average density (of a homogeneous sphere for simplicity) to the dynamical (free-fall) time scale, we may estimate the characteristic radius of the WD as follows: 
\begin{equation}
R_{\rm WD} \sim \left(\frac{3 G M_{\rm WD} t^2}{4 \pi^2}\right)^{1/3} \ ,
\end{equation}
where $G$ is the gravitational constant, and the subscript `WD' refers to the values for the WD remnant. The characteristic expansion velocity here is initially $\sim 100$ km s$^{-1}$ with $R_{\rm WD} \sim 10^{10}$ cm but is decreased to $\sim 10$ km s$^{-1}$ with $R_{\rm WD} \sim 10^{13}$ cm. In a similar manner with the `second' unbound ejecta, we estimate that the diffusion time scale starts exceeding the expansion time scale at $\sim 300$ days when the WD radius has been expanded to $R_{\rm WD} \sim 2 \times 10^{13}$ cm. After this point, most of the deposited energy will be channeled to the radiation energy, and start contributing to the optical output. This component will be very opaque to $\gamma$-rays: 
\begin{equation}
\tau_{\gamma} \sim 4 \times 10^{4} \left(\frac{M_{\rm WD}}{1.33 M_\odot}\right) \left(\frac{R}{2 \times 10^{13} \ {\rm cm}}\right)^{-2} \ , 
\end{equation}
assuming a homogeneous sphere for the remnant WD for simplicity. Therefore, the light curve will follows the $^{56}$Co decay (with full trapping) in this phase.

The analysis of the post-maximum, late-time light curve of SN 2019muj suggests the existence of an inner dense component which keeps opaque for (at least) $\gsim 100$ days \citep[the property that is shared by many SNe Iax;][]{kawabata2021}. In the following, we investigate whether the scenario proposed above, composed of the initial unbound ejecta, the second (delayed) unbound inner component, and the bound WD remnant, can explain the bolometric light curve of SN 2019muj. 

\begin{figure}[t]
\centering
\includegraphics[width=1.0\columnwidth]{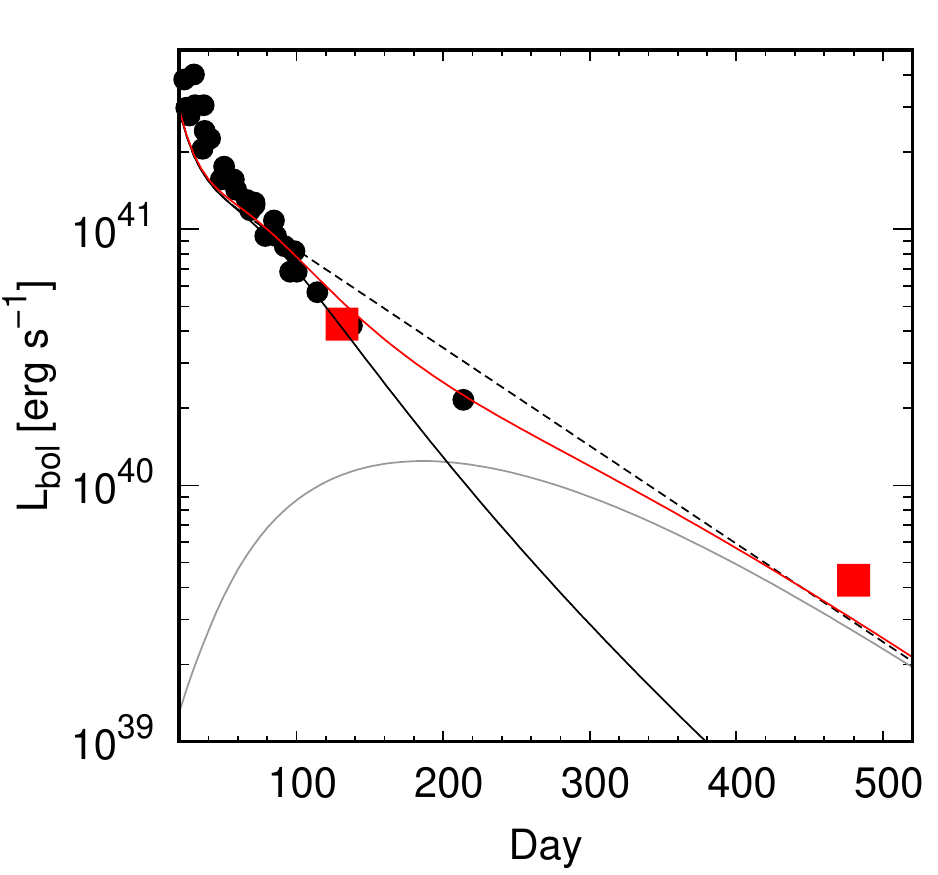}
\caption{The (simplified) late-time light curve model, as compared to the boleometric light curve of SN 2019muj \citep[black squares:][]{kawabata2021} as complemented with the bolometric luminosities adopted in the TARDIS late-time spectral models (red squares). The power deposited to the inner ejecta component is shown by a solid-black line, while the full deposition curve is shown by a dashed-black line. In addition, the contribution by the `innermost' bound WD component is shown by a gray line. The red line shows the combined light curve of the inner `ejecta' and innermost `WD remnant' contributions. 
}
\label{fig:lc}
\end{figure}

Fig. \ref{fig:lc} shows a simplified model where the evolution of the power deposited by the decay $\gamma$-rays (plus positrons) is compared to the bolometric light curve of SN 2019muj. In this figure, the bolometric light curve \citep{kawabata2021} is shifted downward to fit the bolometric luminosity on day 131 obtained through the spectral model presented in this work. Accordingly, in this light curve model, we assume a smaller amount of $^{56}$Ni ($0.015 M_\odot$) in the `second' ejecta than originally suggested by \citet{kawabata2021}. In addition, we have further added an `innermost' bound WD component, with $M$($^{56}$Ni) $= 0.015 M_\odot$, for which the early phase luminosity is truncated with the Gaussian function so that it mimics the diffusion time scale of $\sim 200-300$ days. The outermost component (i.e., the original unbound ejecta) is not included in the model here, whose contribution is largely negligible after $\sim 50$ days since the explosion \citep{kawabata2021}. We note that a possible contribution of the bound WD remnant, based on the model by \citet{shen2017}, was discussed for SN 2012Z with its long-term light curve \citep{mccully2022}. We further note that a possible contribution from a shock-heated He companion star \citep{liu2013,zeng2020} has also been studied for SN 2012Z; this contribution is limited to $\lsim 10^{38}$ erg s$^{-1}$\citep{pan2013} that is well below the luminosity of SN 2019muj on day 480. This contribution may emerge later, which might have provided some contribution in a case of SN 2012Z at $\sim 1,000$ days after the explosion \citep{mccully2022}.

We find that the inner (unbound) component explains the light curve evolution up to $\sim 100 - 150$ days, confirming the suggestion by \citet{kawabata2021}. However, this component alone is not sufficient to explain the later-time evolution after $\sim 200$ days, which roughly follows the $^{56}$Co decay line (or showing even a flatter evolution). Such an extremely long-lasting, slow evolution has indeed been found for a few SNe Iax for which the long-term follow-up observations have been conducted \citep{foley2014,mccully2022}, and it is likely a generic property of SNe Iax. The present work shows that this behavior could be explained by the continuous $^{56}$Co-decay heating within the bound WD remnant, which starts contributing to the optical output after $\sim 1$ year since the explosion. 

The scenario proposed here and the wind-driven outflow scenario by \citet{shen2017} are not mutually exclusive.  It is therefore interesting to see whether the expected structure in the wind model is consistent with the structure derived in the present work. Since the outer `wind' well above the photosphere in the wind model follows the homologous expansion (see above), we extract the density structure of the outflow from Fig. 3 of \citet{shen2017} on day 500, and scale it to days 131 and 480 assuming the homologous expansion. The density structure of the wind model is shown in Fig. \ref{fig:density}. We note that the model here is for a specific combination of $M_{\rm WD} = 0.6 M_\odot$ and $M$($^{56}$Ni) $= 0.03 M_\odot$, and thus the comparison here is for demonstration purpose. Fig. \ref{fig:density} shows that the `wind' has the higher density than the reference model by \citet{barna2021}, and it can be regarded as a distinct inner component. The density is on average lower than the inner component we derived for SN 2019muj. However, under the expected variations in the model predictions by \citet{shen2017} (e.g., the masses of the WD and $^{56}$Ni, and the initial density and temperature structures), we may regard that the match is indeed good, at least in a qualitative level. 
Indeed, the detail of the density structure likely reflects the nature of the ejection process leading to the high-density inner component. As shown in Fig. \ref{fig:density}, the wind-driven outflow model predicts a relatively flat density structure. On the other hand, if the ejection of the remnant WD surface/envelope is more like an instantaneous energy injection, the resulting density structure will be qualitatively similar to the canonical SN ejecta density (with different density and velocity scale), having a flat inner part and a steep outer layer \citep[e.g.,][while it was for a different context.]{moriya2010}. To discriminate between these possibilities, further efforts on (1) deriving the detailed density structure of the inner core for a sample of SNe Iax, and (2) sophisticating the model prediction based on different scenarios, will be necessary. Additionally, the composition structure in the inner component may potentially provide strong diagnostics, as the wind-driven outflow scenario is likely more sensitive to the initial composition structure within the remnant WD and therefore may suffer from some contamination of the unburnt material. Further investigation along this line  will require a higher level of the sophistication than currently employed, both in the spectral modeling and the simulation of the remnant WD evolution.

A remaining question is on the formation of the photosphere. The photospheric radius we have derived is at a few $\times 10^{14}$ cm, and this seems decreasing from day 131 to day 480. A similar behavior is seen in the late-time spectral evolution of SN Iax 2014dt \citep{kawabata2018}. The similarity to SN 2014dt, which are relatively bright for SNe Iax, indicates that the same mechanism is behind SNe Iax in general. These behaviors suggest that the photosphere is still formed within the unbound (second) ejecta (or wind) and has not yet reached to the surface of the remnant WD. The present work predicts that the remnant WD is probably exposed at $\sim 700 - 1,000$ days after the explosion for SN 2019muj. Such extremely late-phase data have been available only for SN 2008ha and 2012Z \citep{foley2014,mccully2022}, and it should be interesting to search for possible signatures of the bound WD remnant in such data. 

\section{Concluding Remarks}\label{sec:summary}

In the present work, we have addressed a possible origin of the peculiar late-phase properties of SNe Iax. By studying the details of the spectral formation processes of the allowed transitions (mainly Fe II and Fe I transitions, and the Ca II NIR) and the forbidden transitions ([Ca II]$\lambda\lambda$7292, 7324 and [O I]$\lambda\lambda$6300, 6363) in the late phase of SN 2019muj, we have shown the existence of a dense inner component (controlling the properties in the late phase) which is distinct from the outer components (controlling the properties in the early phase). Further, we have provided quantitative estimate on the nature of the inner component; the mass of the innermost component is $\sim 0.03 M_\odot$ dominated by Fe (which can be initially $^{56}$Ni; $\sim 0.015 - 0.2 M_\odot$ to be consistent with the light curve evolution), expanding with the velocity of $\sim 760$ km s$^{-1}$. A modest amount of Ca ($\sim 5 \times 10^{-5} M_\odot$) is contained in the inner component, while the C+O-rich material is probably lacking ($\lsim 1.5 \times 10^{-3} M_\odot$). Further completing the picture will require detailed radiation transfer simulations including the self-consistent formation of the photosphere, which is ongoing and will be presented elsewhere. 

The mass of the Fe-rich material we have derived for the inner component is comparable to that within the bound WD remnant predicted in the failed WD thermonuclear explosion scenario, especially the `weakest' end of the model sequence (which is consistent with the early-phase properties of SN 2019muj). This motivates us to consider the scenario in which the inner component was initially a part of the unbound WD but later ejected. We have provided a simplified model in which the $^{56}$Ni-rich materials initially trapped in the bound WD remnant become unbound and ejected (as the `second' ejecta behind the `first' unbound ejecta) by the energy input through the radioactive decay chain of $^{56}$Ni/Co/Fe, which can be connected to the `wind'-driven outflow from the WD remnant \citep{shen2017}. The scenario can largely explain the main characteristics of the inner components we have derived, e.g., the masses of key elements and the ejecta velocity (or kinetic energy), as well as the long-term light curve evolution. Further investigation of the scenario will require radiation hydrodynamic simulations with detailed treatment of physical processes involved \citep[e.g.,][]{shen2017}, coupled with detailed multi-wavelength radiation transfer simulations to provide detailed predictions for observables. We postpone such investigation to the future. 

While we have focused on the properties of SN 2019muj, the present findings have broad implications for the origin of SNe Iax as a class. As we have also shown that the late-time spectra converge to become more and more similar toward the later phases, the existence of the inner component should be a generic feature of SNe Iax. The (late-phase) spectral evolution indeed suggests that the properties of the outer layer (i.e., the `first' unbound ejecta in this scenario) are very diverse among SNe Ia, while those of the inner component (i.e., the `second' ejecta powered by the radioactive input) are similar with each other. On the other hand, the failed deflagration scenario predicts a relation between the properties of the unbound ejecta and those of the bound WD, in a way that more massive, $^{56}$Ni-rich unbound ejecta leave behind a less massive WD remnant. Therefore, whether the proposed scenario can explain properties of SNe Iax, including various similarities and diversities (and correlations between different properties) must be addressed in the future. We are planning to apply similar analyses as presented in this work to a sample of SNe Iax. 

One observational finding which can possibly be a key to understanding the above question is the possible relations between the peak luminosity and some properties in the late-phase spectra. While the sample is still very limited, we find that the properties of late-time spectra may form a sequence set by the peak luminosity. Fainter SNe Iax tend to have the following properties; (1) no or smaller shift in the central wavelength of the [Ca II], (2) stronger [Ca II], and (3) no or weaker forbidden transitions of Fe-peak elements ([Fe II] and [Ni II]), as compared to  brighter SNe Iax. These possible `rules' suggest that the properties of the inner ejecta component (e.g., either in the density or compositions) are linked to the peak luminosity, which may further be related to the nature of the WD remnant at the birth. 

\appendix

\section{Spectra in individual exposure frames} 

\begin{figure*}[t]
\centering
\includegraphics[width=0.49\columnwidth]{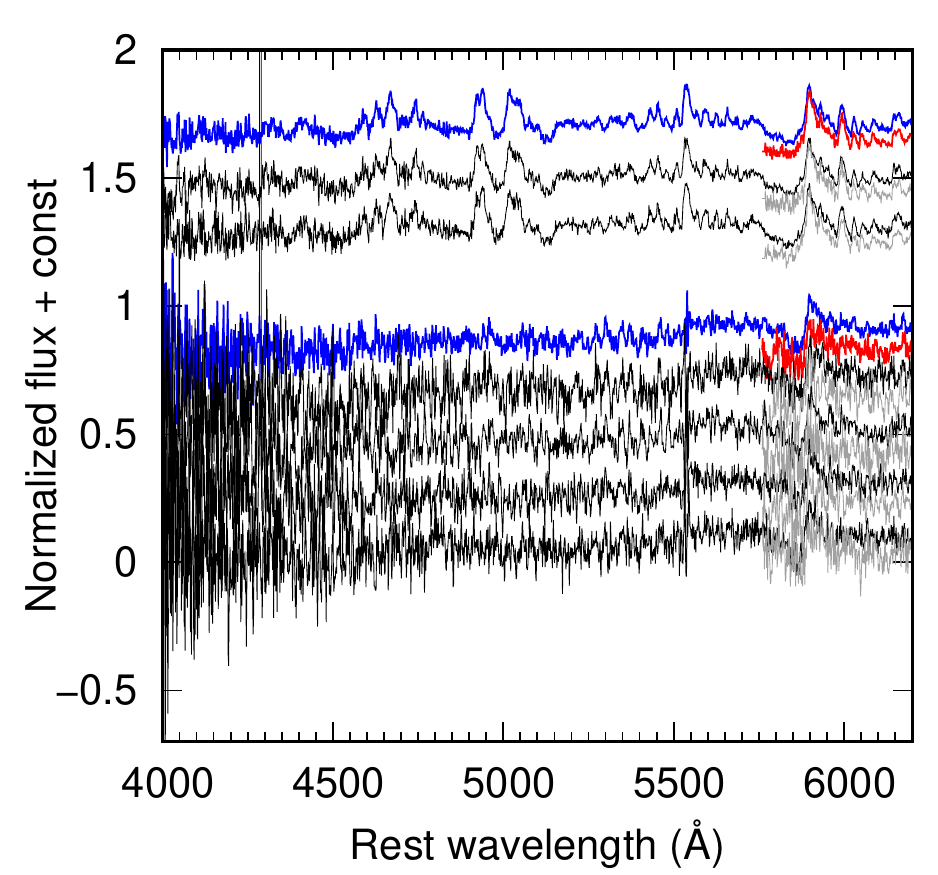}
\includegraphics[width=0.49\columnwidth]{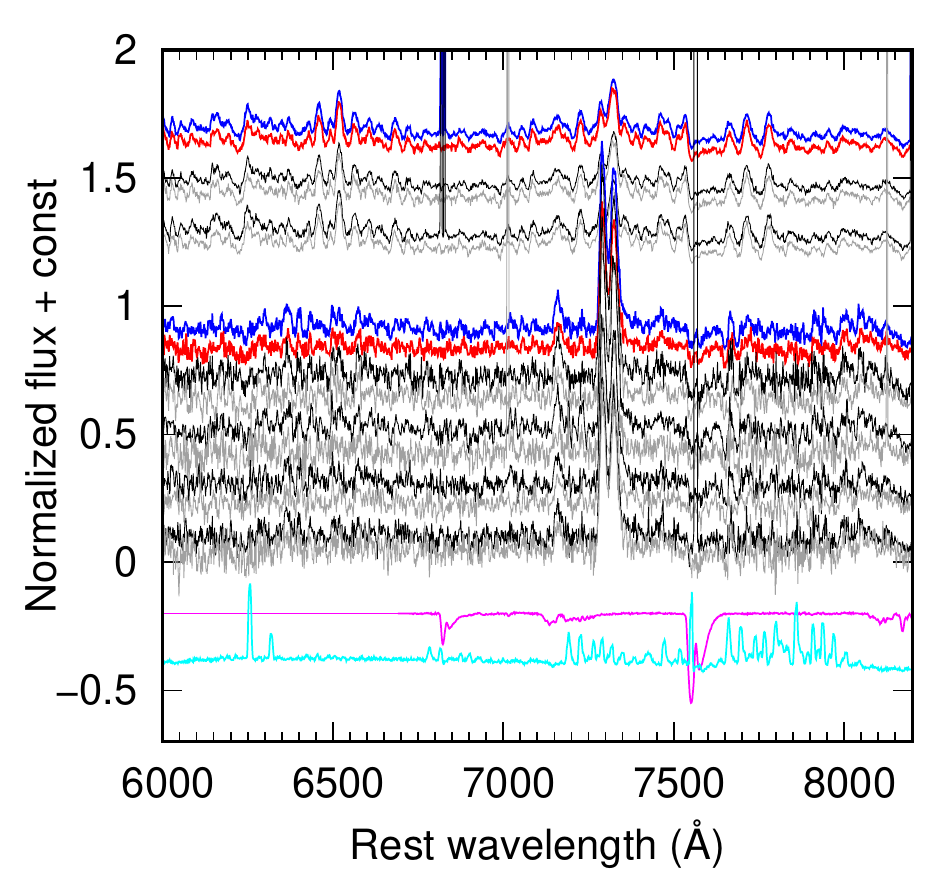}
\includegraphics[width=0.49\columnwidth]{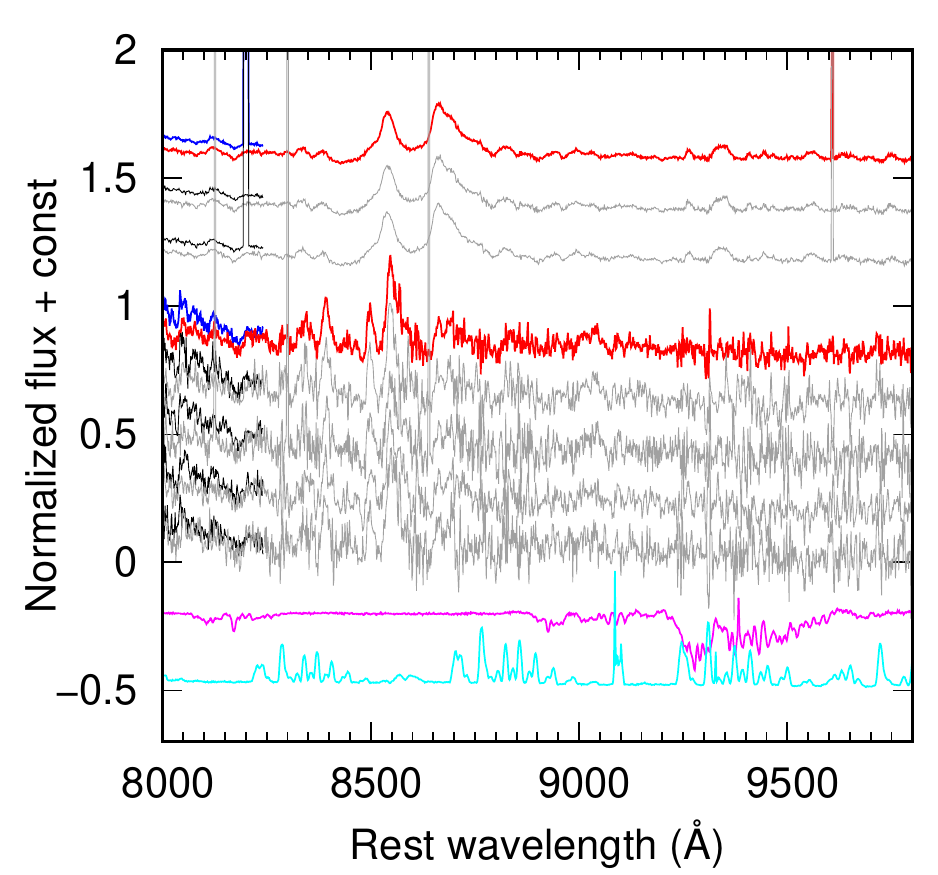}
\caption{The individual frames of the late-time spectra of SN 2019muj. The extinction is not corrected for in this figure. The combined spectra obtained with the B300 (blue) and R300 (red) grisms are shown, as well as the individual frames (black for the B300 and gray for the R300. The top and bottom portions in each panel are for day 131 and day 480, respectively. In addition to the spectra of SN 2019muj, the sky-line spectra (cyan) and the telluric-absorption spectra (magenta) for the R300 grism on day 480 are shown in an arbitrary scale. 
}
\label{fig:spectra_appendix}
\end{figure*}

Fig. \ref{fig:spectra_appendix} shows the spectra of SN 2019muj on days 131 and 480. To check possible artifacts, we have plotted them with the spectra in the individual exposures as well as the sky-emission and telluric-absorption patterns for the red portion of the spectra (as obtained with the R300 grism). There is no doubt that essentially all the spectral-line features seen on day 131 are real, possibly except for the region above $\sim 9,000$\AA\ which might suffer from the telluric absorptions.

The spectrum on day 480 may suffer from artifacts more substantially than the spectrum on day 131. However, we conclude that most of the features are real; (1) strong features are seen in the individual frames and would not be introduced by photon noises, (2) the spectra taken with different setups in the blue and red show good match, (3) there is no clear match between the observed spectral features and the sky/telluric patterns (while these may affect some spectral features at $\gsim 9,000$\AA). 

We further note that (4) many lines seen on day 480 tend to have corresponding features on day 131, and (5) these features are well reproduced by the Fe II transitions. We further note that (6) spectral features match quite well to other SNe Iax in late phases even on day 480 (e.g., see Fig, \ref{fig:o_profile} for the comparison between SNe 2019muj and 2010ae). 

In summary, we conclude that most of the spectral features seen in the late-phase spectra of SN 2019muj as presented here must be real features. A possible exception is the features seen in the red portion of the spectra at $\gsim 9,000$\AA, where strong sky emission lines and telluric absorption might introduce some artifacts.

\acknowledgments

This research is based on data collected at the Subaru Telescope (S19B-055 and S20B-056), which is operated by the National Astronomical Observatory of Japan. We thank Kentaro Aoki and the staff of the Subaru Telescope for their excellent assistance for the observation. We are honored and grateful for the opportunity of observing the Universe from Maunakea, which has the cultural, historical, and natural significance in Hawaii. K.M. acknowledges support from the Japan Society for the Promotion of Science (JSPS) KAKENHI grant JP18H05223, JP20H00174, and JP20H04737. Some data presented in this work are obtained from WISeREP (https://www.wiserep.org). This research made use of TARDIS, a community-developed software package for spectral
synthesis in supernovae. The development of TARDIS received support from GitHub, the Google Summer of Code initiative, and from ESA's Summer of Code in Space program. TARDIS is a fiscally
sponsored project of NumFOCUS. TARDIS makes extensive use of Astropy and Pyne.

%






\bibliography{sn2019muj_focas}{}
\bibliographystyle{aasjournal}



\end{document}